\documentclass[authoryear,preprint,review,8pt]{elsarticle}

\usepackage{natbib}
\setcitestyle{numbers,square}
\usepackage{amssymb}
\usepackage{bm}
\usepackage{amsmath}
\usepackage[lined,boxed,commentsnumbered,ruled]{algorithm2e}
\usepackage{siunitx}
\usepackage{subfigure}
\usepackage{graphicx} 
\usepackage{epstopdf}
\usepackage{booktabs}
\usepackage{xcolor}
\usepackage{changes}
\usepackage{todonotes}
\usepackage{hyperref}

\usepackage{multirow}
\usepackage{booktabs}
\usepackage{array}
\usepackage{amsmath}
\usepackage{siunitx}
\usepackage{xcolor}
\colorlet{red}{black}

\journal{Journal of Computational Physics}

\begin{document}

\begin{frontmatter}



\title{A Discrete Adjoint Gas-Kinetic Scheme for Aerodynamic Shape Optimization in Turbulent Continuum Flows}

 \author[label1]{Hangkong Wu}
 \author[label1]{Yuze Zhu}
 \author[label3]{Yajun Zhu}
 \author[label1,label4,label5]{Kun Xu\corref{cor1}}
\cortext[cor1]{Corresponding author}
 \ead{makxu@ust.hk}
\affiliation[label1]{organization={Department of Mathematics, Hong Kong University of Science and Technology},
             city={ Clear Water Bay, Kowloon},
             state={Hong Kong},
             country={China}}
 \affiliation[label3]{organization={Research and Development Office, Shanghai Suochen Information Technology Co., Ltd},
             city={Shanghai},
             country={China}}
 \affiliation[label4]{organization={Department of Mechanical and Aerospace Engineering, Hong Kong University of Science and Technology},
             city={ Clear Water Bay, Kowloon},
             state={Hong Kong},
             country={China}}
\affiliation[label5]{organization={Shenzhen Research Institute, Hong Kong University of Science and Technology},
             city={Shenzhen},
             country={China}}

\begin{abstract}

This study presents an efficient and accurate discrete adjoint gas-kinetic scheme (GKS) for sensitivity analysis and aerodynamic shape optimization in continuum flow regimes. Developed using the backward mode of algorithmic differentiation (AD), the adjoint solver is rigorously verified against a duality-preserving linearized GKS solver generated via forward-mode AD. The robustness and practical effectiveness of the solver are evaluated through three benchmark cases: the inverse design of turbine blades, lift-to-drag ratio enhancement, and shock-strength reduction for a NACA 0012 airfoil. To capture realistic flow physics, fully turbulent optimizations are conducted using the one-equation Spalart–Allmaras (SA) model. Numerical results demonstrate excellent agreement between the discrete adjoint and linearized solvers, exhibiting matching sensitivity convergence behaviors, identical asymptotic residual decay rates, and negligible discrepancies in final sensitivity predictions. Furthermore, the optimization studies confirm that targeted design objectives are consistently achieved within a limited number of design cycles, highlighting the solver’s computational efficiency, accuracy, and suitability for complex aerodynamic geometries.

\end{abstract}

\begin{keyword}
discrete adjoint, gas-kinetic scheme, sensitivity evaluation, shape optimization, algorithmic differentiation
\end{keyword}

\end{frontmatter}

\section{Introduction}\label{introduction}

Improving aerodynamic performance is essential for reducing operating costs in commercial aviation, making shape optimization a cornerstone of modern aircraft design. Driven by advancements in numerical algorithms and high-performance computing, computational fluid dynamics (CFD)-based optimization has garnered significant interest~\cite{HAN2024101007}. Existing CFD-driven approaches generally fall into two categories: gradient-free and gradient-based methods. While gradient-free strategies—often coupled with surrogate models~\cite{2024HZHKring}—excel at global design space exploration, their computational costs scale rapidly with the number of design variables, limiting their viability for high-dimensional problems. Conversely, gradient-based methods, particularly those utilizing the adjoint approach for sensitivity evaluation, offer superior computational efficiency~\cite{LI2026110842,XU2015175}. Because the primal and adjoint systems are solved only once per design cycle, the overall computational expense remains largely independent of the number of design parameters. This independence enables the effective optimization of complex geometries across expansive design spaces.

\begin{figure}[h!]
    \centering
    \includegraphics[width=0.9\linewidth]{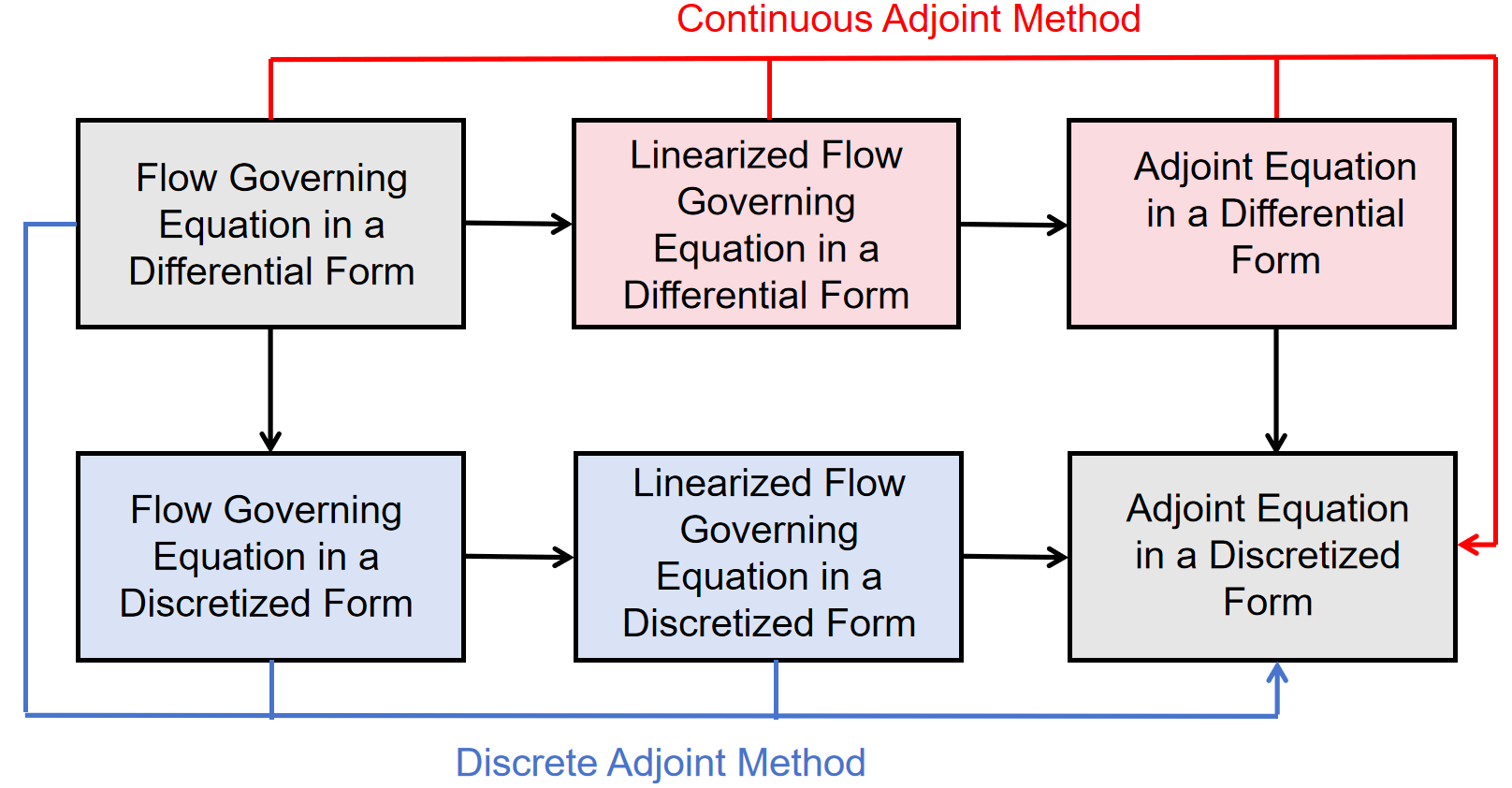}
    \caption{The flow chart of different forms of adjoint methods}
    \label{adj_type}
\end{figure}

The adjoint method was first introduced to the computational fluid dynamics (CFD) community by Jameson \textit{et al}.\cite{Jameson1988AerodynamicDV}, who derived the governing adjoint equations and boundary conditions using control theory\cite{Jameson1998}. Subsequent advancements have extended these techniques to multistage turbomachinery~\cite{Vitale2020,aerospace10020106}, unsteady flows~\cite{Ma2016,RUBINO2020106132,Wu2024JoT}, and multidisciplinary design environments~\cite{He2010,Duta2003,Wu2023JoT,Wu2024AIAA}. Today, two principal variations of the method are widely utilized: the continuous and discrete adjoint approaches (see Fig.\ref{adj_type}). The discrete adjoint formulation ensures exact consistency with the numerical discretization of the primal solver, typically yielding more accurate sensitivities and enhanced robustness\cite{Nadarajah2000}. Furthermore, the advent of algorithmic differentiation (AD) tools~\cite{Tapenade} has made the development of fully turbulent discrete adjoint solvers significantly more tractable~\cite{Giles2003}. In contrast, manually linearizing turbulence model equations within the continuous adjoint framework remains highly challenging, frequently necessitating the assumption of constant eddy viscosity~\cite{Wang2010}. AD-based implementations generally entail differentiating the primal solver’s subroutines and assembling them to construct a complete adjoint system. The complexity of developing a discrete adjoint solver is therefore heavily dependent on the structure of the numerical flux employed in the primal flow solver. In conventional upwind finite-volume schemes~\cite{ROE1997250}, inviscid fluxes are evaluated using exact or approximate Riemann solvers, whereas viscous fluxes are typically discretized separately via central differencing. This decoupled treatment introduces structural inconsistencies that complicate the formulation of the adjoint system, particularly concerning the viscous terms.

The gas-kinetic scheme (GKS)\cite{XU19949,XU2001289} circumvents this issue by deriving numerical fluxes from the integral solution of the Bhatnagar–Gross–Krook (BGK) model\cite{LI2006532}, which yields a time-dependent distribution function at the cell interface. Macroscopic fluxes are subsequently evaluated by taking the velocity moments of this distribution. Because the distribution function inherently encapsulates both equilibrium and non-equilibrium states, the inviscid and viscous fluxes are computed simultaneously. This unified treatment significantly reduces the complexity associated with adjoint solver development. Furthermore, the kinetic flux formulation provides enhanced shock-capturing robustness~\cite{ZHAO2022110812,JXAIAA} and facilitates compact, high-order discretizations in both space and time~\cite{PAN2016197,JI2018446}. Over the past two decades, the GKS has been successfully applied to a diverse array of flow problems, consistently demonstrating exceptional robustness and accuracy.

Recent studies have also explored adjoint-based design optimization within kinetic-theory frameworks. Notably, Yuan and Wu~\cite{YUAN2024113366,YUAN2025114102} proposed a hybrid adjoint formulation wherein the adjoint equations are solved continuously, while the sensitivities are evaluated discretely. Through a series of design optimizations spanning continuum to rarefied flow regimes, they demonstrated the efficacy of adjoint-based kinetic approaches. The primal solvers employed in their research are based on the Boltzmann equation and are solved via the discrete velocity method. However, compared to the GKS, direct Boltzmann solvers typically exhibit slower convergence due to the stiff collision term, necessitating specialized numerical treatments to maintain efficiency in continuum flow regimes~\cite{YUAN2020106972}. Additionally, their optimization studies focused primarily on laminar Navier–Stokes (NS) flows, omitting fully turbulent optimizations that incorporate turbulence model equations. Because the manual linearization of turbulence models is notoriously challenging within continuous-adjoint frameworks, extending these methods to fully turbulent regimes is difficult. Consequently, adjoint-based shape optimization for continuum turbulent flows within the GKS framework remains largely unexplored, presenting a critical area for further investigation.

In this work, a fully turbulent discrete adjoint gas-kinetic solver is developed to facilitate consistent and efficient sensitivity analysis for aerodynamic shape optimization. To accurately model turbulent flow physics, the one-equation Spalart–Allmaras (SA) model~\cite{SA1992} is employed for eddy-viscosity evaluation. The discrete adjoint solver is implemented utilizing the backward mode of algorithmic differentiation (AD) and is rigorously verified against a duality-preserving linearized GKS solver constructed via forward-mode AD. This comprehensive verification process encompasses detailed comparisons of the flow and adjoint fields, sensitivity convergence behaviors, and asymptotic root-mean-square (RMS) residual decay rates between the two solvers. Finally, the practical capability of the proposed solver is demonstrated through three benchmark optimization problems: the inverse design of turbine blades, lift-to-drag ratio enhancement, and shock-strength reduction for a NACA 0012 airfoil.

\section{Methodology}\label{methodology}
In this section, the numerical formulation of the GKS and the corresponding linearized and discrete‑adjoint formulations are presented in detail.

\subsection{Gas-Kinetic Scheme}\label{gks_sec}
The two-dimensional gas-kinetic equation based on the BGK model is given by
\begin{equation}\label{GKS}
\frac{\partial f}{\partial t} + u\frac{\partial f}{\partial x} + v\frac{\partial f}{\partial y}=\frac{g-f}{\tau}
\end{equation}
where $f$ denotes the distribution function, ($u$,$v$) are the microscopic particle velocities, ($x$,$y$) are the spatial grid coordinates, $\tau$ is the collision time, and $g$ is the distribution function at the equilibrium state, satisfying the following Maxwellian distribution
\begin{equation}
g = \rho\Big(\frac{\lambda}{\pi}\Big)^{\frac{K+2}{2}}e^{-\lambda[(u-U)^2 + (v-V)^2 +\xi^2]}
\end{equation}
where ($U$,$V$) are the macroscopic velocity components, $\rho$ is the density, $\lambda=\frac{\rho}{2p}$ with $p$ denoting the static pressure, $K$ is the number of internal degrees of freedom (equal to 3 for a diatomic gas), and $\xi_i(i=1,2,\dots,K)$ represents the internal variable for each degree of freedom, with $\xi^2=\xi_1^2 + \xi_2^2 +\dots+\xi_K^2$.

Due to the conservation law during particle collisions, the compatibility condition of the collision term in Eq.~\ref{GKS} is given by
\begin{equation}\label{compat}
\int\frac{g-f}{\tau}\Psi d\Xi=0
\end{equation}
where $\Psi$ is the collision invariant and has the following form
\begin{equation}\label{invariant}
\Psi = [\Psi_1, \Psi_2, \Psi_3, \Psi_4]^T=[1, u, v, \frac{1}{2}(u^2+v^2+\xi^2)]^T
\end{equation}

The relationships between the macroscopic flow variables and fluxes and the distribution function are obtained through the velocity‑space moments, given by
\begin{align}
Q   &= \int f\Psi   d\Xi\label{flowvar}\\
\boldsymbol{F} &= \int f\Psi \boldsymbol{u} d\Xi\label{Flux}
\end{align}
where $\boldsymbol u=(u,v)$, $\boldsymbol F = (F_x,F_y)$, and $d\Xi=dudvd\xi$.

In the above formulae, the distribution function $f$ is obtained by the integral solution of Eq.~\ref{GKS}.
\begin{equation}\label{int_f}
f(\boldsymbol x,\boldsymbol u,\xi,t)=\frac{1}{\tau}\int_{0}^{t} g({\boldsymbol x}^{'},\boldsymbol u,\xi,t^{'})e^{{-(t-t^{'})}/\tau_n} dt^{'} + e^{-\frac{t}{\tau_n}}f_0(\boldsymbol x_0,\boldsymbol u,\xi,0)
\end{equation}
where $\boldsymbol x=(x,y)$, ${\boldsymbol x}^{'}=\boldsymbol x-\boldsymbol u(t-t^{'})$ is the particle trajectory, $\boldsymbol x_0=x-ut$ is the initial position of a particle, $f_0$ is the initial distribution function at the beginning of each time step, $\tau$ and $\tau_n$ are the physical and numerical collision time, respectively. The numerical collision time is introduced into the integral solution to enhance numerical dissipation, thereby improving shock‑capturing capability. For fully turbulent flows, they are determined by
\begin{gather}
\begin{aligned}
\tau &= \frac{\mu_L + \mu_T}{p}\\
\tau_n &=\frac{\mu_L}{p}+C|\frac{p_l-p_r}{p_l+p_r}|\Delta t
\end{aligned}
\end{gather}
where $p_l$ and $p_r$ are the pressure at the left and right hand sides of a cell interface, $\mu_L$ and $\mu_T$ are the laminar and eddy viscosity coefficients where $\mu_T$ is determined by the SA turbulence model and $\mu_L$ is computed by the Sutherland's law, $\Delta t$ is the time step, and $C$ is a constant, typically chosen in the range $1\leq C \leq 10$. For notational simplicity, we set $\boldsymbol x=0$ in the following derivations.

According to the Chapman-Enskog (C-E) expansion, $f_0$ in Eq.~\ref{int_f} at the cell interface can be expressed by
\begin{gather}\label{exp_f0}
\begin{aligned}
f_0(\boldsymbol x_0,\boldsymbol u,\xi,0) =& g^{l}(\boldsymbol 0,\boldsymbol u,\xi,0)[1-(t+\tau){\boldsymbol a}^{l}\cdot \boldsymbol u-\tau{\boldsymbol A}^{l}]H[u_n]+\\
&g^{r}(\boldsymbol 0,\boldsymbol u,\xi,0)[1-(t+\tau){\boldsymbol a}^{r}\cdot \boldsymbol u-\tau{\boldsymbol A}^{r}][1-H[u_n]]
\end{aligned}
\end{gather}
where $g^{l}$ and $g^{r}$ are the equilibrium distribution at the left and right hand sides of an interface, and are determined by the reconstructed macroscopic flow variables $Q^l$ and $Q^r$, $u_n$ is the normal part of $\boldsymbol u$, $H[u_n]$ is the Heaviside function, $\boldsymbol {a}^l$ and ${\boldsymbol a}^r$ are the spatial derivatives of $g^l$ and $g^r$, while $\boldsymbol A^l$ and $\boldsymbol A^r$ are the temporal derivatives of $g^l$ and $g^r$.

According to the Taylor expansion, $g$ in Eq.~\ref{int_f} can be expressed by
\begin{equation}\label{exp_g}
g(\boldsymbol x^{'},\boldsymbol u, \xi, t^{'}) = g^c(\boldsymbol 0,\boldsymbol u,\xi,0)(1-\boldsymbol a^c\cdot \boldsymbol u(t-t^{'}) + \boldsymbol A^c t^{'} )
\end{equation}
where $g^c$ is the equilibrium distribution function computed by the kinetic-averaged macroscopic flow variables $Q^c$ at the cell interface.

In Eqs.~\ref{exp_f0} and ~\ref{exp_g}, $\boldsymbol a^{l/r/c}$ and $\boldsymbol A^{l/r/c}$ are computed using the following formulae
\begin{gather}
\begin{aligned}
\boldsymbol a^{l/r/c}=\frac{1}{g^{l/r/c}}\frac{\partial g^{l/r/c}}{\partial x}=\frac{\partial ln g^{l/r/c}}{\partial x}\\
\boldsymbol A^{l/r/c}=\frac{1}{g^{l/r/c}}\frac{\partial g^{l/r/c}}{\partial t}=\frac{\partial ln g^{l/r/c}}{\partial t}
\end{aligned}
\end{gather}
To save space, the detailed expressions of $\boldsymbol a^{l/r/c}$ and $\boldsymbol A^{l/r/c}$ will be omitted here; interested readers can refer to Ref. \text{\cite{XU2001289}} for the full derivations.

Substituting Eqs.~\ref{exp_f0} and ~\ref{exp_g} into Eq.~\ref{int_f}, the distribution function $f$ at $\boldsymbol x=0$ can be expressed by
\begin{gather}\label{df}
\begin{aligned}
f(\boldsymbol 0,\boldsymbol u,\xi,t)=&(1-e^{-\frac{t}{\tau_n}})g^c+\\
&[(t+\tau)e^{-\frac{t}{\tau_n}}-\tau]\boldsymbol a^c\cdot \boldsymbol ug^c +\\
&(t-\tau+\tau e^{-\frac{t}{\tau_n}})\boldsymbol A^c g^c + \\
&e^{-\frac{t}{\tau_n}}[g^l H[u_n] + g^r(1-H[u_n])]-\\
&e^{-\frac{t}{\tau_n}}(\tau + t)[\boldsymbol a^l\cdot \boldsymbol u g^l H[u_n] + \boldsymbol a^r\cdot \boldsymbol u g^r (1-H[u_n])] - \\
&\tau e^{-\frac{t}{\tau_n}}[g^l\boldsymbol A^l H[u_n] + g^r\boldsymbol A^r(1-H[u_n])]
\end{aligned}
\end{gather}

Substituting Eq.~\ref{df} into Eqs.~\ref{flowvar} and ~\ref{Flux}, we can obtain the macroscopic fluxes and flow variables at the cell interface.

\subsection{Linearized/Adjoint Principles}

To simplify the subsequent derivations, the optimization problem based on the gas‑kinetic scheme can be expressed in the following symbolic form:
\begin{align}
&min: I = I(Q_i,Q_b,\boldsymbol x)\label{objfun}\\
&s.t.1: R(f,Q_i,Q_b,\mu_T,\boldsymbol x)=0 \label{st1}\\
&s.t.2: f = f(Q_i,Q_b,\boldsymbol x)\label{st2} \\
&s.t.3: Q_b= Q_b(Q_i,\boldsymbol x)\label{st3}\\
&s.t.4: \mu_T = \mu_T(Q_i,Q_b,\boldsymbol x)\label{st4}
\end{align}
where Eq.~\ref{objfun} defines the objective function, and Eqs.~\ref{st1}$\sim$~\ref{st4} represent the four constraints. The quantities $Q_i$ and $Q_b$ denote the macroscopic flow variables in the interior computational domain and at the dummy cells, respectively. The objective function $I$ in Eq.~\ref{objfun} may represent isentropic efficiency for internal flows or lift-to-drag ratio for external flows, and is determined by the flow variables and the grid coordinates.

The first constraint corresponds to the flow governing equation. The residual vector $R$ in Eq.~\ref{st1} depends on the distribution function, the flow variables, the grid coordinates, and the eddy viscosity coefficient.
When conventional upwind schemes are used for residual evaluation, the residual must be decomposed into inviscid and viscous parts (see Ref.~\cite{Wu2021}) and treated separately, which increases the complexity of developing the discrete adjoint solver.
As discussed in Sec.~\ref{gks_sec}, the distribution function $f$ (also referred to as the second constraint) can be expressed directly in terms of the flow variables and the grid coordinates, as shown in Eq.~\ref{st2}. The third constraint corresponds to the boundary conditions imposed in the flow GKS solver. The final constraint is associated with the turbulence model equation adopted in this work.

For gradient-based design optimization, the sensitivities of $I$ with respect to $\boldsymbol x$ are required. Applying the chain rule, the linearization of Eq.~\ref{objfun} yields
\begin{equation}\label{dIda}
\frac{dI}{d\boldsymbol x}=\frac{\partial I}{\partial Q_i}\frac{d Q_i}{d\boldsymbol x} + \frac{\partial I}{\partial Q_b}\frac{d Q_b}{d\boldsymbol x}+\frac{\partial I}{\partial \boldsymbol x}
\end{equation}

In Eq.~\ref{dIda}, the terms $\frac{\partial I}{\partial Q_i}$, $\frac{\partial I}{\partial Q_b}$, and $\frac{\partial I}{\partial \boldsymbol x}$ do not require solving the flow governing equations, and may be evaluated either numerically or analytically. In contrast, the remaining terms in Eq.~\ref{dIda} rely on the linearized GKS solver, and the manner in which these derivatives are computed largely determines the overall efficiency of the optimization procedure.

First, linearizing Eq.~\ref{st1} yields
\begin{equation}\label{dRda}
\frac{dR}{d\boldsymbol x} = \frac{\partial R}{\partial f}\frac{df}{d\boldsymbol x}  + \frac{\partial R}{\partial Q_i}\frac{d Q_i}{d\boldsymbol x} + \frac{\partial R}{\partial Q_b}\frac{d Q_b}{d\boldsymbol x} + \frac{\partial R}{\partial \mu_T}\frac{d \mu_T}{d\boldsymbol x} + \frac{\partial R}{\partial \boldsymbol x}=0
\end{equation}

To evaluate $\frac{df}{d\boldsymbol x}$,  $\frac{dQ_b}{d\boldsymbol x}$, and $\frac{d\mu_T}{d\boldsymbol x}$, it is necessary to linearize Eqs.~\ref{st2}$\sim$\ref{st4}
\begin{align}
\frac{df}{d\boldsymbol x}&=\frac{\partial f}{\partial Q_i}\frac{dQ_i}{d\boldsymbol x}+\frac{\partial f}{\partial Q_b}\frac{dQ_b}{d\boldsymbol x}+\frac{\partial f}{\partial \boldsymbol x}\label{dfda}\\
\frac{dQ_b}{d\boldsymbol x}&=\frac{\partial Q_b}{\partial Q_i}\frac{dQ_i}{d\boldsymbol x}+\frac{\partial Q_b}{\partial \boldsymbol x}\label{dQbda}\\
\frac{d\mu_T}{d\boldsymbol x}&=\frac{\partial \mu_T}{\partial Q_i}\frac{dQ_i}{d\boldsymbol x}+\frac{\partial \mu_T}{\partial Q_b}\frac{dQ_b}{d\boldsymbol x}+\frac{\partial \mu_T}{\partial \boldsymbol x}\label{dmuda}
\end{align}

Second, substituting Eqs.~\ref{dfda}$\sim$\ref{dmuda} into Eq.~\ref{dRda} yields
\begin{equation}\label{lin_eqn}
\frac{dR}{d\boldsymbol x}=A\frac{d Q_i}{d\boldsymbol x} + B=0
\end{equation}
where $A$ is the Jacobian matrix and $B$ is a column vector. Their detailed expressions are given by
\begin{align}
A &= \frac{\partial R}{\partial f}\Big(\frac{\partial f}{\partial Q_i} + \frac{\partial f}{\partial Q_b}\frac{\partial Q_b}{\partial Q_i}\Big) + \frac{\partial R}{\partial \mu_T}\Big(\frac{\partial \mu_T}{\partial Q_i} + \frac{\partial \mu_T}{\partial Q_b}\frac{\partial Q_b}{\partial Q_i}\Big)+\frac{\partial R}{\partial Q_b}\frac{\partial Q_b}{\partial Q_i} + \frac{\partial R}{\partial Q_i}\\
B &= \frac{\partial R}{\partial f}\Big(\frac{\partial f}{\partial \boldsymbol x} + \frac{\partial f}{\partial Q_b}\frac{\partial Q_b}{\partial \boldsymbol x}\Big) + \frac{\partial R}{\partial \mu_T}\Big(\frac{\partial \mu_T}{\partial \boldsymbol x} + \frac{\partial \mu_T}{\partial Q_b}\frac{\partial Q_b}{\partial \boldsymbol x}\Big)+\frac{\partial R}{\partial Q_b}\frac{\partial Q_b}{\partial \boldsymbol x} + \frac{\partial R}{\partial \boldsymbol x}
\end{align}

For a linearized GKS solver, it first solves Eq.~\ref{lin_eqn} to obtain $\frac{dQ_i}{d\boldsymbol x}$, which is then substituted into Eq.~\ref{dIda} to compute sensitivities. However, the number of CFD evaluations required to compute $\frac{dQ_i}{d\boldsymbol x}$ scales with the number of design variables. When the number of design variables is far more than the number of objective functions, the linearized approach suffers from low computational efficiency.

Unlike the linearized approach, the adjoint method reorganizes the sequence of operations to avoid explicitly solving Eq.~\ref{lin_eqn}. First, obtain the explicit expression of $\frac{dQ_i}{d\boldsymbol x}$ in Eq.~\ref{lin_eqn} and then substitute it into Eq.~\ref{dIda}, yielding
\begin{equation}\label{dIda_adj}
\frac{dI}{d\boldsymbol x}=-\lambda^T B+\frac{\partial I}{\partial \boldsymbol x} + \frac{\partial I}{\partial Q_b}\frac{\partial Q_b}{\partial \boldsymbol x}
\end{equation}
where $\lambda^T$ is the adjoint variable vector and satisfies the following adjoint equation
\begin{equation}\label{adj_eqn}
A^T\lambda = (\frac{\partial I}{\partial Q_i}+\frac{\partial I}{\partial Q_b}\frac{\partial Q_b}{\partial Q_i})^T
\end{equation}

When the adjoint method is used to compute sensitivities, two steps are required: (1) solve the adjoint equation (Eq.~\ref{adj_eqn}) to obtain the adjoint variables $\lambda$; and (2) substitute $\lambda$ into Eq.~\ref{dIda_adj} to evaluate sensitivities. Since the adjoint equation is independent of the number of design variables, both the primal and adjoint equations need to be solved only once to evaluate sensitivity per design cycle. Consequently, when hundreds or even thousands of design variables are used for shape parameterization, the adjoint method is significantly more efficient than the linearized approach. Furthermore, because the Jacobian matrix in the adjoint system corresponds to the transpose of those in the linearized system, they possess identical eigenvalues. According to the duality‑preserving principle~\cite{Giles2003}, the adjoint and linearized solvers therefore exhibit the same asymptotic residual convergence behavior and identical sensitivity convergence characteristics.

\section{Algorithmic Differentiation}

Algorithmic differentiation provides two operational modes—forward and backward.
In the development of sensitivity solvers, forward‑mode AD is typically used to develop the linearized solver, whereas backward‑mode AD is used to develop the adjoint solver.
To clarify their distinction, we begin by considering the following application of the chain rule:
\begin{equation}
\boldsymbol x \to Q(\boldsymbol x) \to I(Q(\boldsymbol x))
\end{equation}
where $\boldsymbol x$ is the independent variable, $Q$ is the intermediate variable, and $I$ is the dependent variable.

\subsection{Forward Mode}
In forward mode, differentiation begins by perturbing the independent variable $\boldsymbol x$. The resulting perturbations of intermediate variables and ultimately the dependent variables are then propagated using the chain rule:
\begin{equation}\label{forward}
\boldsymbol x_d \to Q_d = \frac{\partial Q}{\partial \boldsymbol x}\boldsymbol x_d \to I_d = \frac{\partial I}{\partial Q}Q_d = \frac{\partial I}{\partial Q}\frac{\partial Q}{\partial \boldsymbol x}\boldsymbol x_d = \frac{\partial I}{\partial \boldsymbol x}\boldsymbol x_d
\end{equation}
where the subscript $d$ represents the forward mode. In the above expression, assigning $\boldsymbol x_d=1$ gives $I_d$ which corresponds to the sensitivity of $I$ with respect to $\boldsymbol x$.

\subsection{Backward Mode}

In the backward mode, sensitivity propagation is performed in the reverse computational direction: the dependent variables are perturbed first, followed by the intermediate and independent variables. The resulting expressions are
\begin{equation}\label{backward}
I_a \to Q_a = \Big(\frac{\partial I}{\partial Q}\Big)^T I_a \to \boldsymbol x_a = \Big(\frac{\partial Q}{\partial \boldsymbol x}\Big)^T Q_a = \Big(\frac{\partial I}{\partial Q}\frac{\partial Q}{\partial \boldsymbol x}\Big)^T I_a = \Big(\frac{\partial I}{\partial \boldsymbol x}\Big)^T I_a
\end{equation}
where the subscript $a$ represents the backward mode. By assigning $I_a=1$, the computed $\boldsymbol x_a$ gives the sensitivity information.

By comparing Eq.~\ref{backward} with Eq.~\ref{forward}, one can see that both modes yield identical sensitivity information, even though their computational sequences differ.

\section{AD‑Based Solver Development}

The development of a discrete adjoint solver using AD is strongly dependent on the structure of the underlying flow solver. Therefore, we first provide a detailed description of the flow solver, including its subroutines and the functionality of each component, before introducing the development of the discrete adjoint solver. The development of the linearized solver is presented separately in Appendix~A.

\subsection{Flow GKS Solver}

Figure~\ref{Flow_GKS} shows the data flow of the flow GKS solver, which consists of seven main subroutines. The functionality of each part is given by
\begin{enumerate}
\item INIT: initialize macroscopic variables $Q_i$ and $Q_b$;
\item BC: apply boundary conditions;
\item DF: compute the distribution function $f$ (Eq.~\ref{df});
\item SA: compute the eddy viscosity using the SA turbulence model;
\item RES: evaluate the residual $R$ in each control volume;
\item UPDATE: update flow variables using the time-marching method;
\item OBJ: compute the objective function after convergence.
\end{enumerate}

Note that the variables in parentheses in Fig.~\ref{Flow_GKS} represent the output on the left‑hand side of the equation and the input on the right‑hand side.

\begin{figure}[h!]
    \centering
\includegraphics[width=0.4\linewidth]{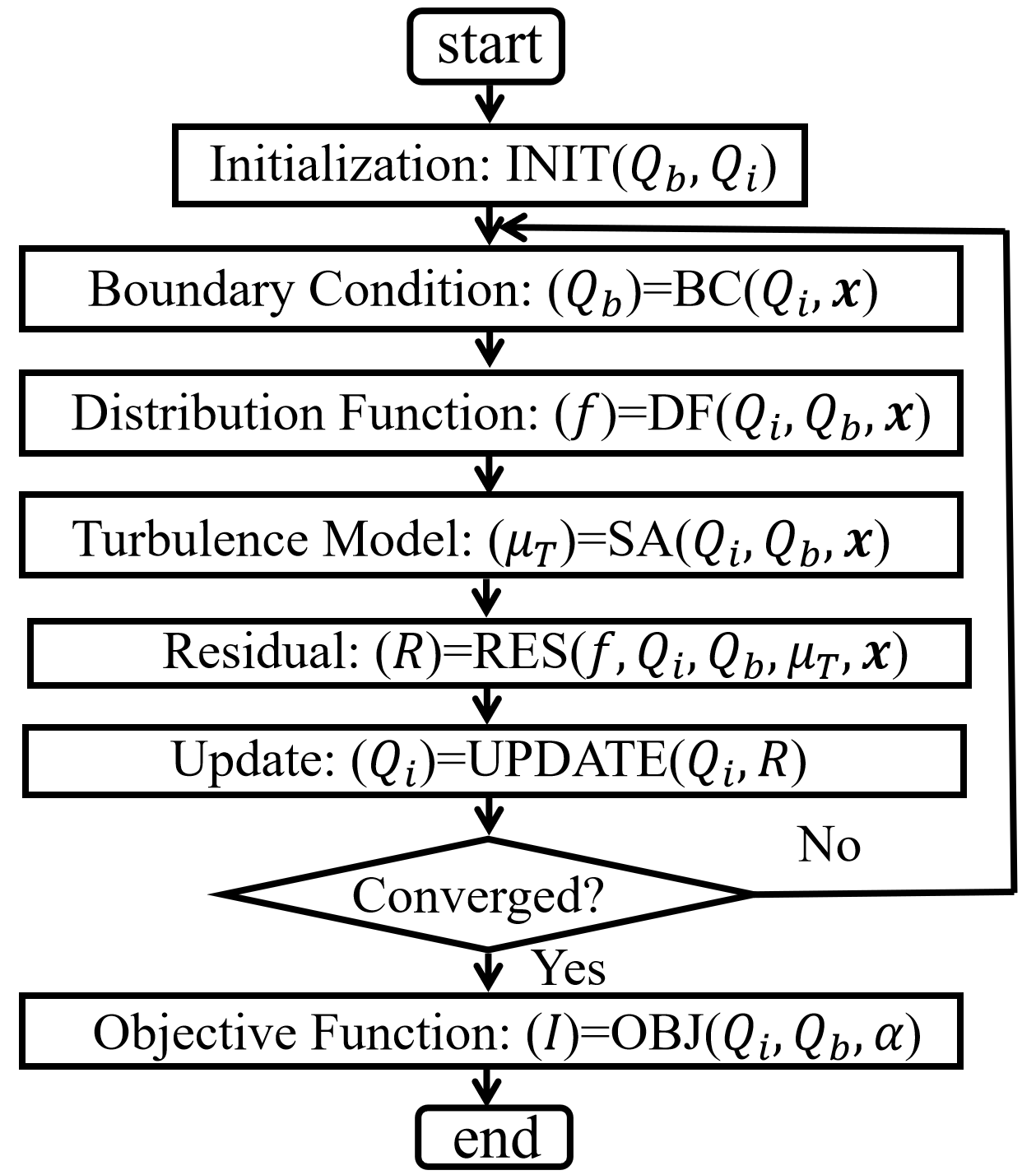}
    \caption{The dataflow of the flow GKS solver}
    \label{Flow_GKS}
\end{figure}

\subsection{Adjoint GKS Solver}

\begin{figure}[h!]
    \centering
\includegraphics[width=0.6\linewidth]{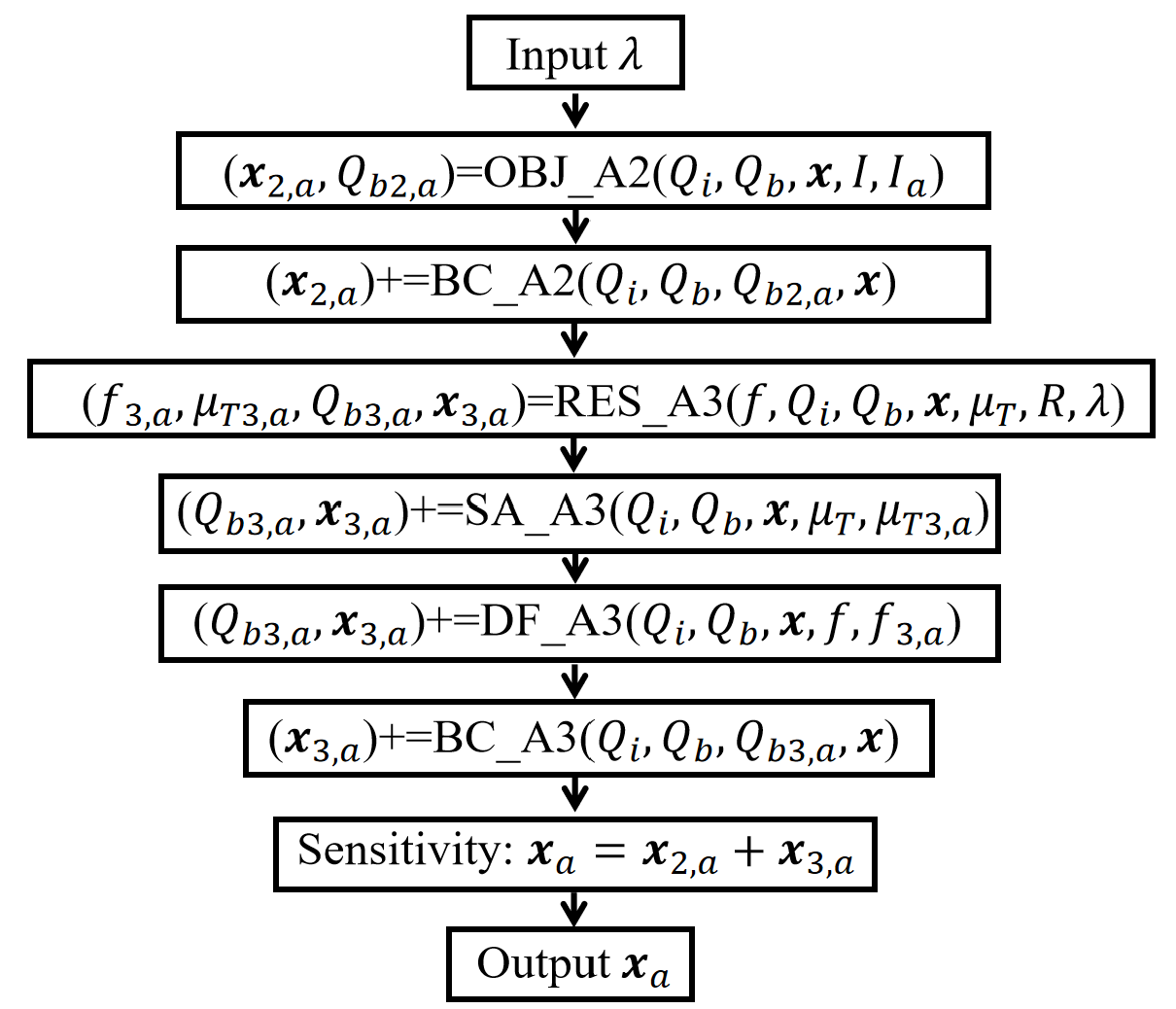}
    \caption{The dataflow of the second part of the adjoint GKS solver: sensitivity evaluation}
    \label{ADJ_GKS2}
\end{figure}

The adjoint GKS solver contains two parts: adjoint equation as shown in Fig.~\ref{ADJ_GKS}, and sensitivity evaluation as shown in Fig.~\ref{ADJ_GKS2}. Since the discrete solver is developed using the backward mode of AD, the assembly of the differentiated codes must likewise proceed in a reverse order.

\subsubsection{Adjoint GKS Equation}
Because the adjoint equation is independent of the design variables, the design variables are kept frozen during the differentiation process at this stage.

\begin{figure}[h!]
    \centering
\includegraphics[width=0.6\linewidth]{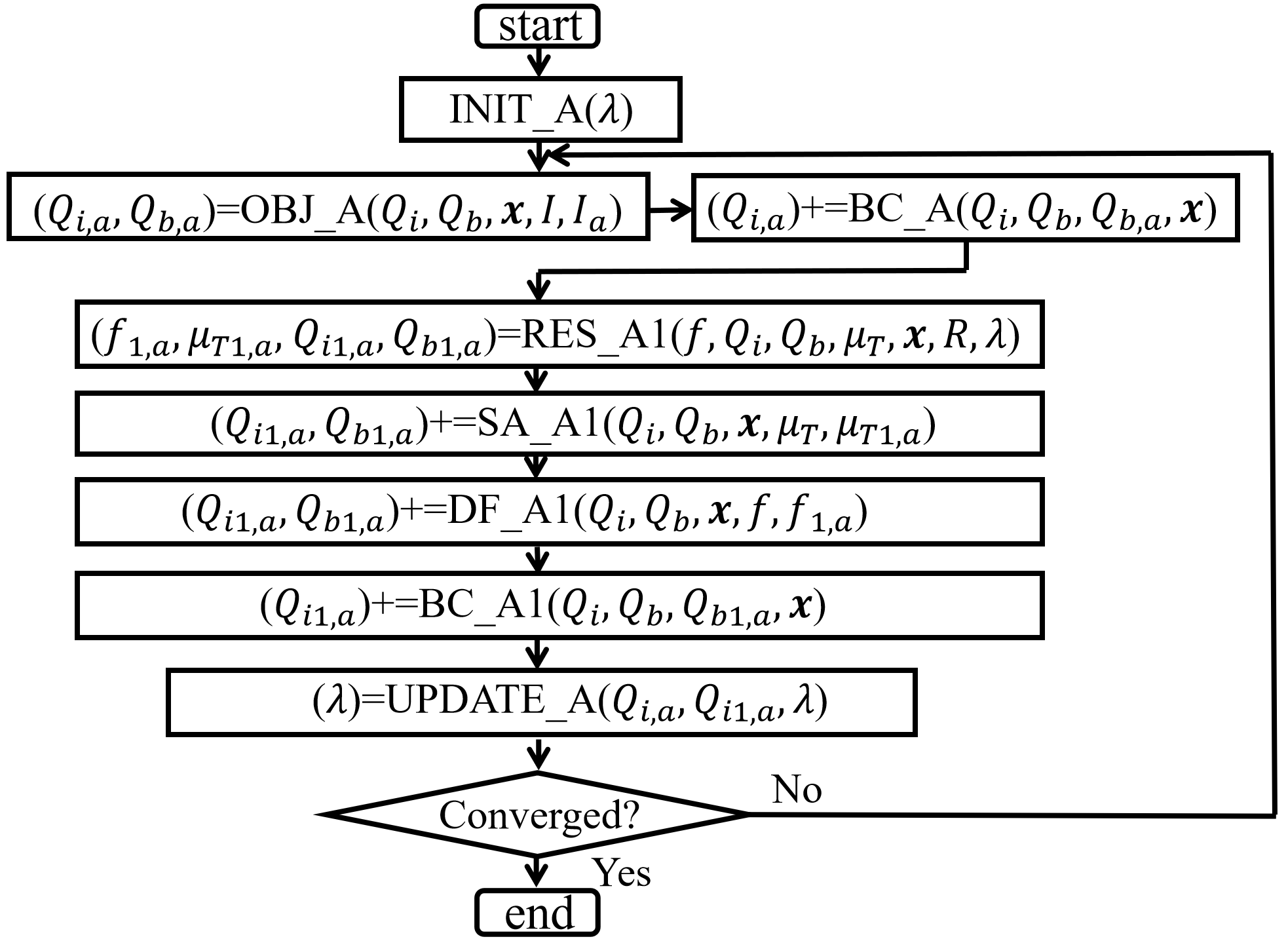}
    \caption{The dataflow of the first part of the adjoint GKS solver: adjoint GKS equation}
    \label{ADJ_GKS}
\end{figure}

\begin{enumerate}
\item INIT\_A: initialize the adjoint variables $\lambda$ with 0;

\item OBJ\_A: differentiate the  objective function to compute the partial derivatives of $I$ with respect to $Q_i$ and $Q_b$;
\begin{equation}\label{step1}
Q_{i,a} = \Big(\frac{\partial I}{\partial Q_i}\Big)^T I_a, Q_{b,a} = \Big( \frac{\partial I}{\partial Q_b}\Big)^T I_a
\end{equation}
where $I_a$ is set to 1.

\item BC\_A: differentiate the boundary condition to transfer the gradient information stored in $Q_{b,a}$ to $Q_{i,a}$, and we have
\begin{equation}\label{step2}
Q_{i,a} = Q_{i,a} + \Big(\frac{\partial Q_b}{\partial Q_i}\Big)^T Q_{b,a}
\end{equation}

Note that $+=$ in Fig.~\ref{ADJ_GKS} represents accumulation.
Substituting Eq.~\ref{step1} into ~\ref{step2}, $Q_{i,a}$ is identifies as the right-hand side of the adjoint GKS equation(see Eq.~\ref{adj_eqn}).

\item RES\_A1: differentiate the residual;
\begin{gather}\label{step3}
\begin{aligned}
&f_{1,a}=\Big(\frac{\partial R}{\partial f}\Big)^T R_a,
\mu_{T1,a}=\Big(\frac{\partial R}{\partial \mu_T}\Big)^T R_a
\\
&Q_{i1,a}=\Big(\frac{\partial R}{\partial Q_i}\Big)^T R_a,
Q_{b1,a}=\Big(\frac{\partial R}{\partial Q_b}\Big)^T R_a
\end{aligned}
\end{gather}
where $R_a$ denotes the adjoint variable $\lambda$.

\item SA\_A1: differentiate the turbulence model equation;
\begin{gather}\label{step4}
\begin{aligned}
Q_{i1,a} &= Q_{i1,a} + \Big(\frac{\partial \mu_T}{\partial Q_i}\Big)^T \mu_{T1,a}\\
Q_{b1,a} &= Q_{b1,a} + \Big(\frac{\partial \mu_T}{\partial Q_b}\Big)^T \mu_{T1,a}\\
\end{aligned}
\end{gather}

\item DF\_A1: differentiate the distribution function;
\begin{gather}\label{step5}
\begin{aligned}
Q_{i1,a} &= Q_{i1,a} + \Big(\frac{\partial f}{\partial Q_i}\Big)^T f_{1,a}\\
Q_{b1,a} &= Q_{b1,a} + \Big(\frac{\partial f}{\partial Q_b}\Big)^T f_{1,a}\\
\end{aligned}
\end{gather}

\item BC\_A1: differentiate the boundary condition;
\begin{equation}\label{step6}
Q_{i1,a} = Q_{i1,a} + \Big(\frac{\partial Q_b}{\partial Q_i}\Big)^T Q_{b1,a}
\end{equation}

By substituting Eqs.~\ref{step4}$\sim$~\ref{step6} into Eq.~\ref{step3}, $Q_{i1,a}$ is found to equal $A^T\lambda$, which corresponds to the left-hand side of the adjoint equation (see Eq.~\ref{adj_eqn}).

\item UPDATE\_A: use the same time-marching scheme as the flow GKS solver to update the adjoint variables.
\begin{equation}
\frac{\partial \lambda}{\partial t} + Q_{i1,a}=Q_{i,a}
\end{equation}

\end{enumerate}

\subsubsection{Sensitivity Evaluation}

In the sensitivity evaluation stage, $\boldsymbol x$ and $Q_b$ are differentiated, while $Q_i$ is frozen. The adjoint variables $\lambda$ serve as the inputs.

\begin{enumerate}

\item OBJ\_A2: differentiate the objective function;
\begin{equation}\label{step1_2}
\boldsymbol x_{2,a} = \Big(\frac{\partial I}{\partial \boldsymbol x}\Big)^T I_a,
Q_{b2,a} = \Big( \frac{\partial I}{\partial Q_b}\Big)^T I_a
\end{equation}
where $I_a$ is set to 1.

\item BC\_A2: differentiate the boundary condition;
\begin{equation}\label{step2_2}
\boldsymbol x_{2,a} = \boldsymbol x_{2,a} + \Big(\frac{\partial Q_b}{\partial \boldsymbol x}\Big)^T Q_{b2,a}
\end{equation}

Substituting Eq.~\ref{step1_2} into ~\ref{step2_2} shows that $\boldsymbol x_{2,a}$ in Eq.~\ref{step2_2} equals the second and third terms in the right-hand side of Eq.~\ref{dIda_adj}.

\item RES\_A3: differentiate the residual;
\begin{gather}\label{step3_2}
f_{3,a}=\Big(\frac{\partial R}{\partial f}\Big)^T \lambda,
\mu_{T3,a}=\Big(\frac{\partial R}{\partial \mu_T}\Big)^T \lambda,
\\
Q_{b3,a}=\Big(\frac{\partial R}{\partial Q_b}\Big)^T\lambda,
\boldsymbol x_{3,a}=\Big(\frac{\partial R}{\partial \boldsymbol x}\Big)^T\lambda
\end{gather}

\item SA\_A3: differentiate the turbulence model equation;
\begin{gather}\label{step4_2}
\begin{aligned}
Q_{b3,a} &= Q_{b3,a} + \Big(\frac{\partial \mu_T}{\partial Q_b}\Big)^T \mu_{T3,a}\\
\boldsymbol x_{3,a} &= \boldsymbol x_{3,a} + \Big(\frac{\partial \mu_T}{\partial \boldsymbol x}\Big)^T \mu_{T3,a}\\
\end{aligned}
\end{gather}

\item DF\_A3: differentiate the distribution function;
\begin{gather}\label{step5_2}
\begin{aligned}
Q_{b3,a} = Q_{b3,a} + \Big(\frac{\partial f}{\partial Q_b}\Big)^T f_{3,a}\\
\boldsymbol x_{3,a} = \boldsymbol x_{3,a} + \Big(\frac{\partial f}{\partial \boldsymbol x}\Big)^T f_{3,a}\\
\end{aligned}
\end{gather}

\item BC\_A3: differentiate the boundary condition;
\begin{equation}\label{step6_2}
\boldsymbol x_{3,a}=\boldsymbol x_{3,a} + \Big(\frac{\partial Q_b}{\partial \boldsymbol x}\Big)^T Q_{b3,a}
\end{equation}

By substituting Eqs.~\ref{step3_2}$\sim$~\ref{step5_2} into ~\ref{step6_2}, $\boldsymbol x_{3,a}$ is equal to the first term in the right-hand side of Eq.~\ref{dIda_adj}

\item Sensitivity: compute the sensitivity.
\begin{equation}\label{step7_2}
\boldsymbol x_a = \boldsymbol x_{2,a} + \boldsymbol x_{3,a}
\end{equation}

$\boldsymbol x_a$ in Eq.~\ref{step7_2} is the calculated sensitivity.

\end{enumerate}

\section{Solver Verification \& Validation}

The flow GKS solver is validated using the experimental data of the Durham turbine cascade case and the NACA 0012 airfoil. The adjoint GKS solver is verified using the linearized GKS solver based on the duality-preserving principle~\cite{Giles2003}.

\subsection{Flow GKS Solver Validation}




\subsubsection{Durham Turbine Cascade Case}

\begin{figure}[h!]
    \centering
\includegraphics[width=0.4\linewidth]{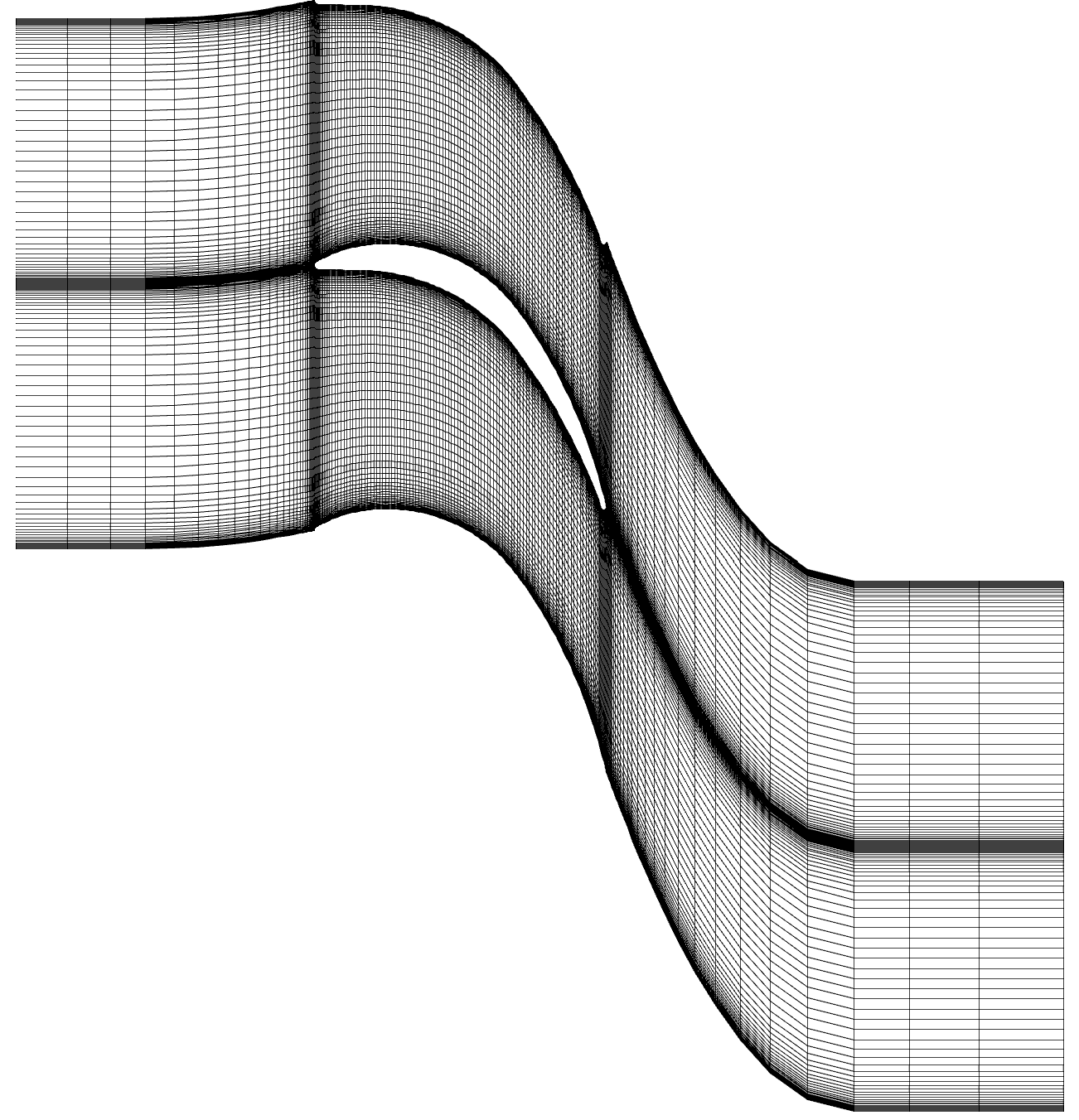}
    \caption{The computational mesh for the Durham turbine cascade case}
    \label{durham_mesh}
\end{figure}

Figure \ref{durham_mesh} presents the computational mesh for the Durham turbine cascade case. The mesh has a density of 145$\times$65 in the x and y directions, including 129 grid points on the blade surfaces.
The boundary conditions are specified as follows. At both the inlet and outlet boundaries, the subsonic non-reflective boundary conditions are imposed. At the inlet boundary, the total pressure of $107882$Pa, the total temperature of $293$K, and the flow angle of $0^{\circ}$ are specified. At the outlet boundary, the area-averaged back pressure of $101325$Pa is prescribed. A Reynolds number of $2.2\times10^5$, based on the blade chord and the isentropic exit velocity, is used. On the solid walls, the adiabatic non-slip wall boundary condition is applied. Periodic boundary condition is imposed on the geometric periodic boundaries.

\begin{figure}[h!]
\centering
\subfigure[whole computational domain]{
	\includegraphics[width=2.0in]{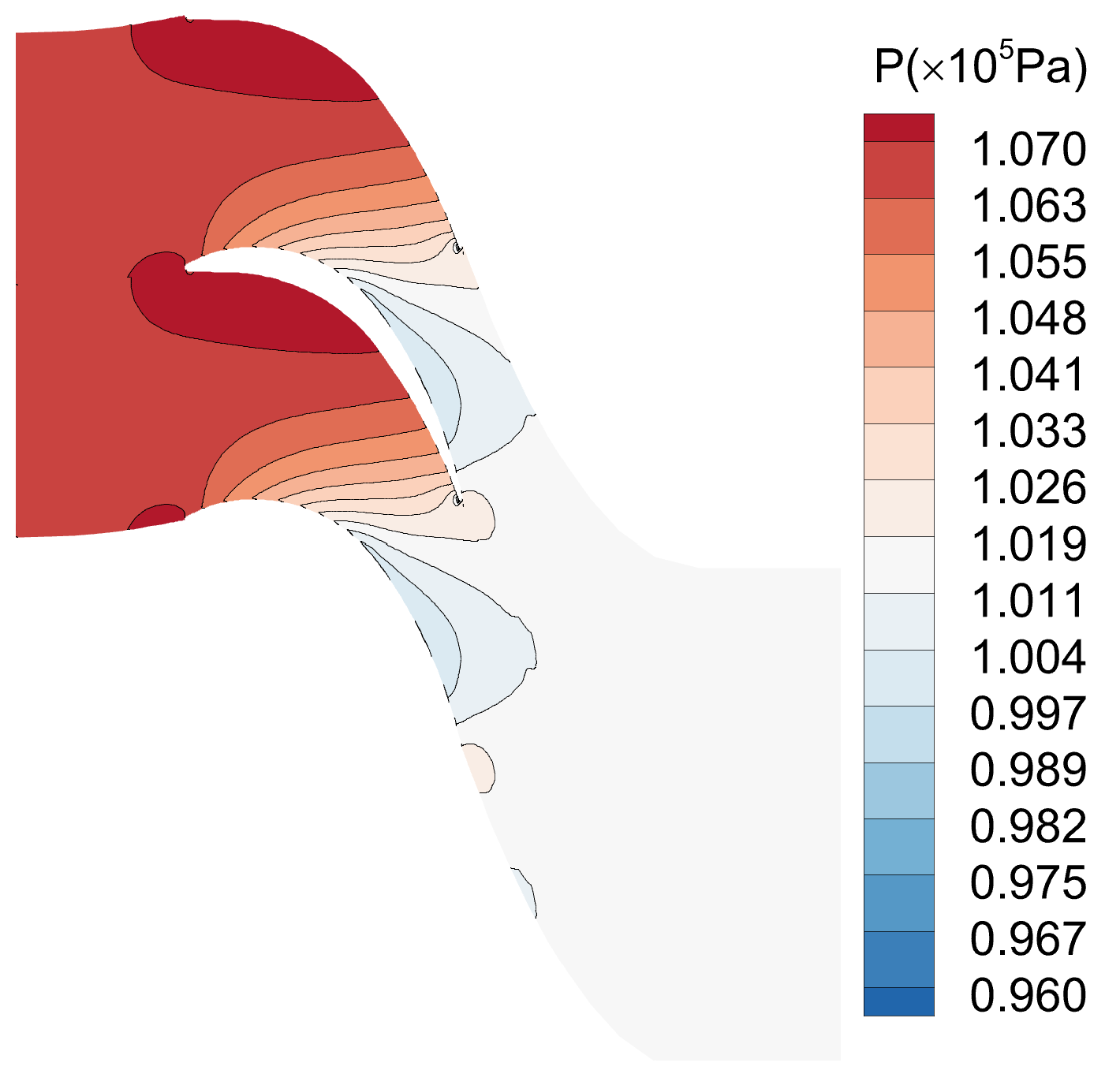}
 \label{Cp_turbine0}
 }
 \subfigure[blade surface]{
	\includegraphics[width=2.0in]{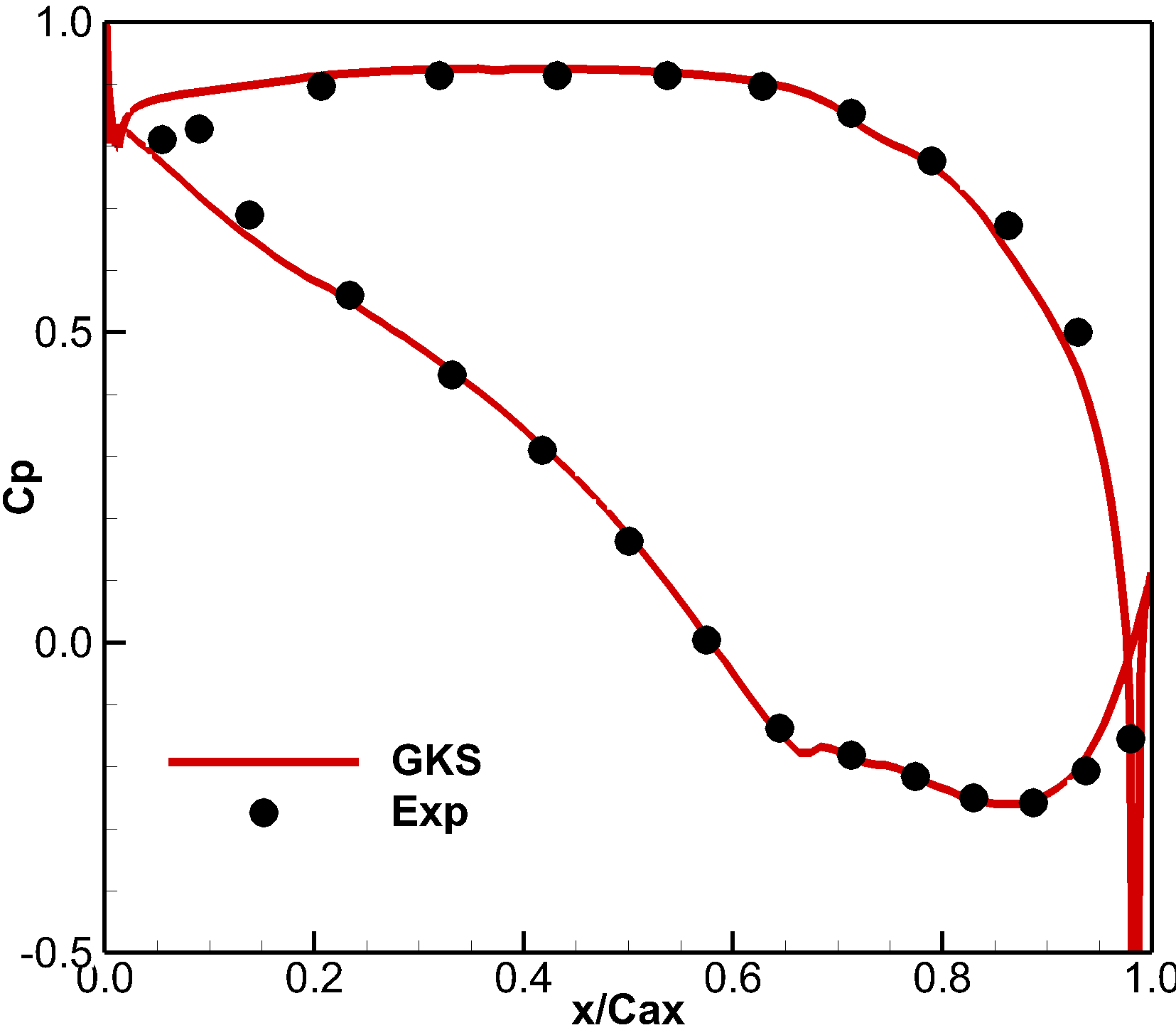}
 \label{Cp_turbine}
 }
	\caption{Distributions of $p$ and $C_p$ for the Durham turbine cascade case: a) in the whole computational domain; b) on blade surfaces (Exp in the legend represents the experimental data)}
\end{figure}

To compare the numerical results to the experimental data, the pressure coefficient ($C_p$) on the blade surface is defined by
\begin{equation}
C_p = \frac{p-p_b}{p_0-p_b}
\end{equation}
where $p_b$ is the back pressure, $p_0$ is the inlet total pressure and $p$ is the static pressure on the blade surface.
Figure \ref{Cp_turbine0} shows the pressure coefficient contours in the whole computational domain and Figure \ref{Cp_turbine} compares the pressure coefficient on the blade surfaces between the numerical results and the experimental data. It can be seen that the two solutions agree well with each other.

\subsubsection{NACA 0012 Airfoil}

\begin{figure}[h!]
    \centering
\includegraphics[width=0.5\linewidth]{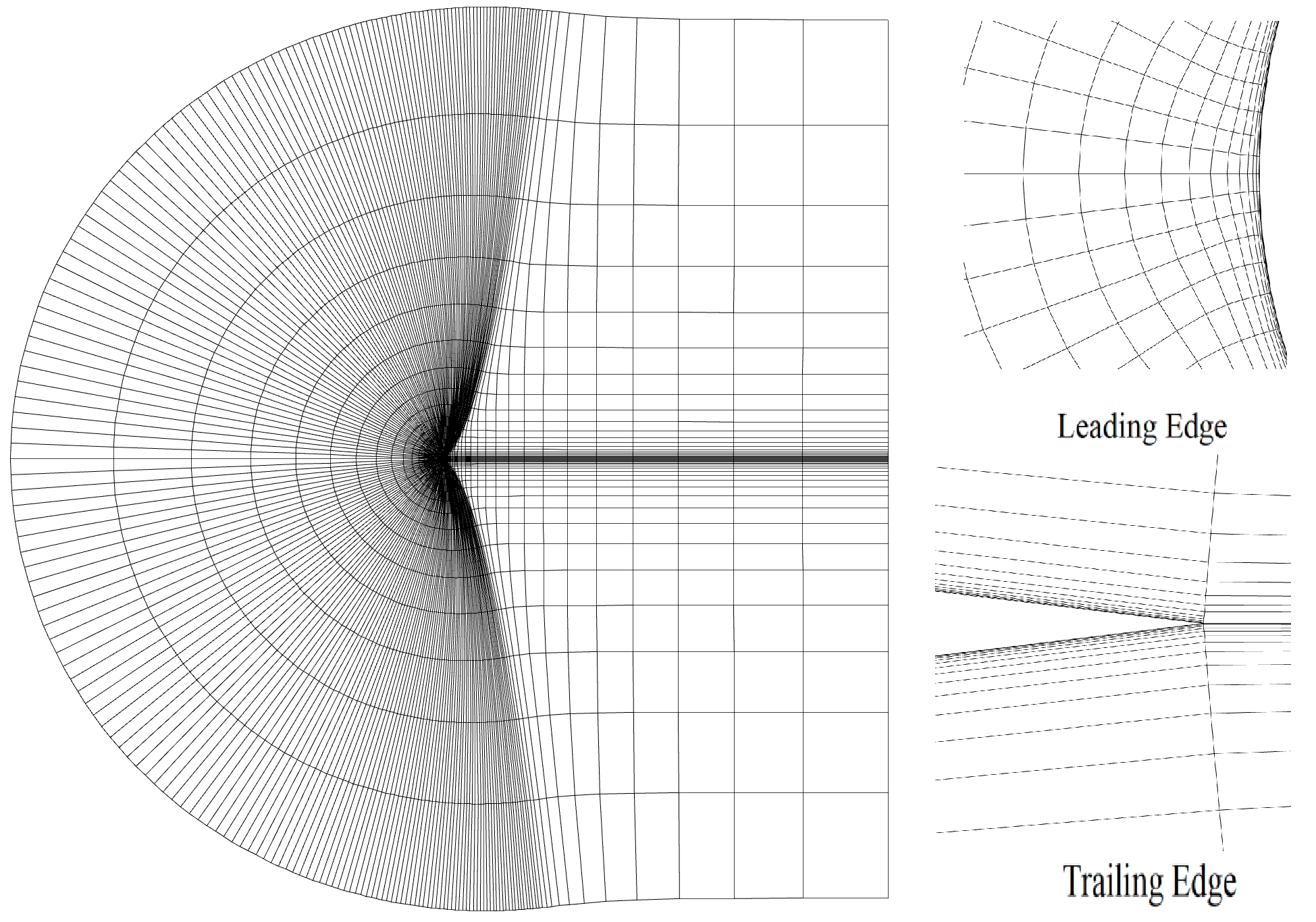}
    \caption{The computational mesh of the NACA 0012 airfoil}
    \label{NACA_mesh}
\end{figure}

The second validation case for the
flow GKS solver is the NACA 0012 airfoil. Figure \ref{NACA_mesh} shows the C-type computational mesh, which has a resolution of 225$\times$65 and includes 129 grid points on the airfoil surface. The farfield boundary is located approximately 500 chord lengths away. The farfield parameters are Mach number of 0.15, Reynolds number based on the chord is 6 million, the angle of attack of 0$^{\circ}$, temperature of 300K, and pressure of 101325Pa. Adiabatic no-slip wall boundary condition is applied on the airfoil surface. For comparison, the experimental data of Gregory et al. is used as reference. The pressure coefficient on the airfoil surface is defined by
\begin{equation}
C_p = \frac{p-p_\infty}{
\frac{1}{2}\rho_\infty U_\infty^2}
\end{equation}
where $\rho_{\infty}$ is the farfield density, $U_{\infty}$ is the farfield velocity.

\begin{figure}[h!]
\centering
\subfigure[whole computational domain]{
	\includegraphics[width=2.0in]{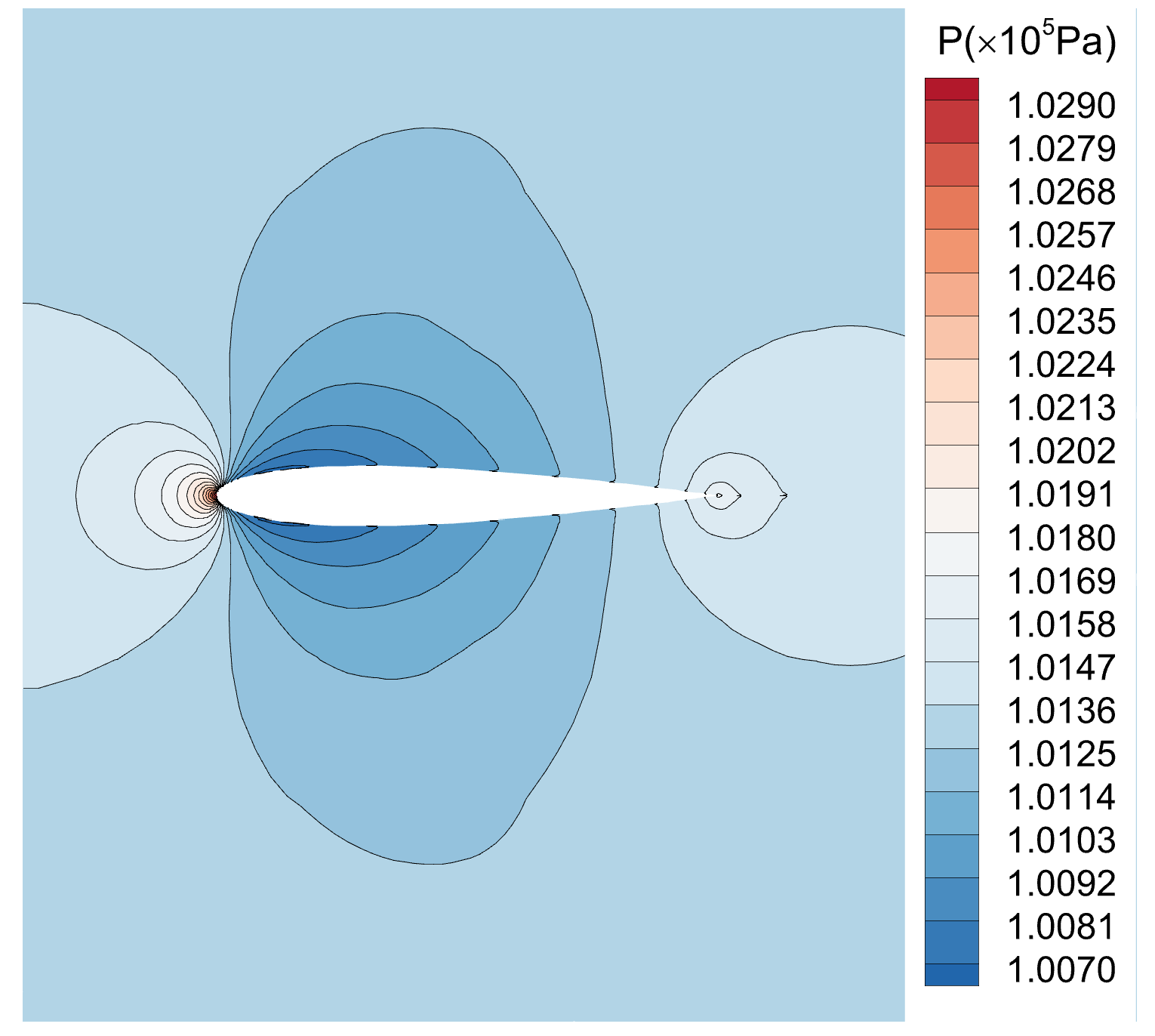}
 \label{NACA_cp0}
 }
 \subfigure[airfoil surface]{
	\includegraphics[width=2.0in]{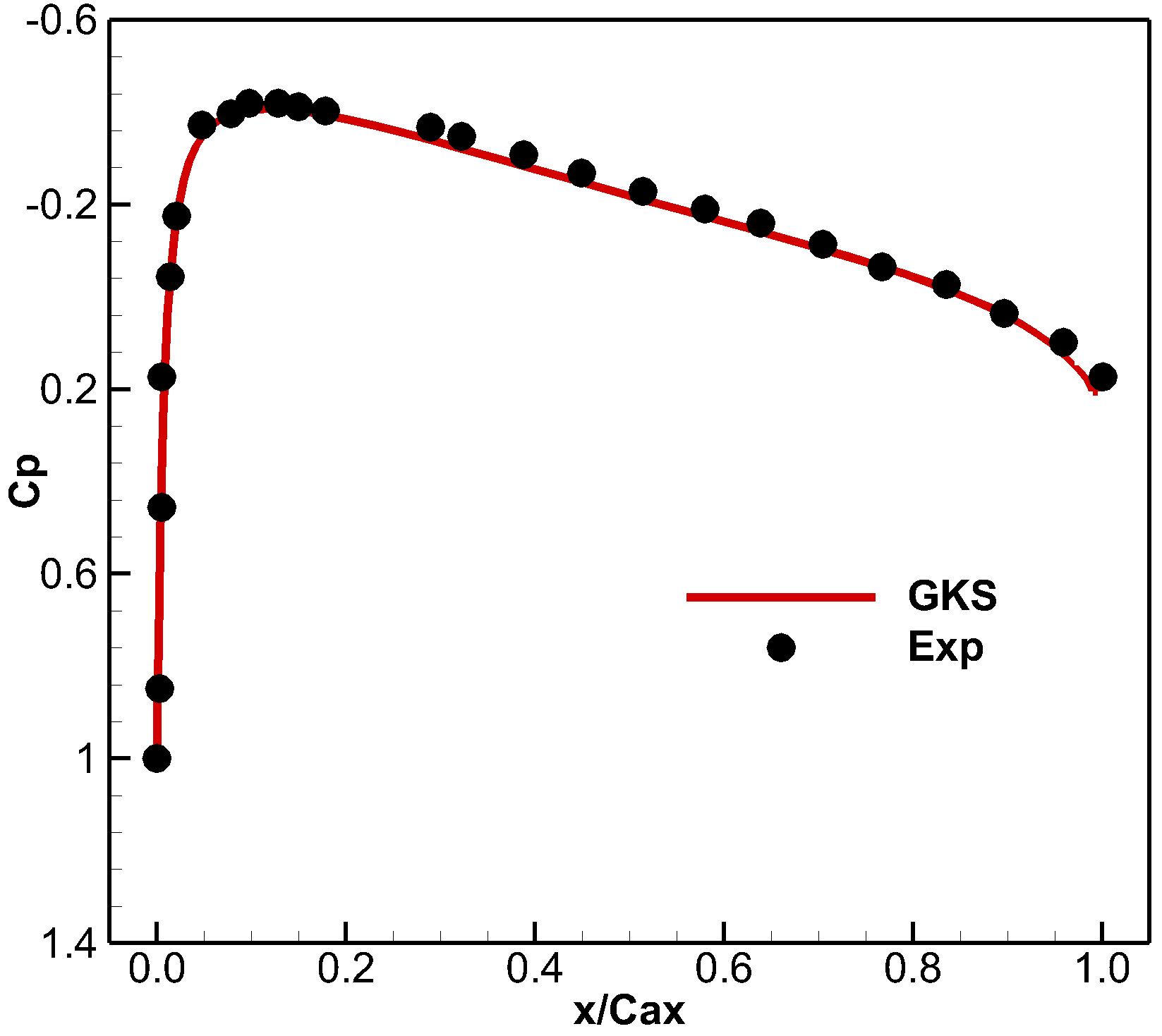}
 \label{NACA_cp}
 }
	\caption{The distributions of $p$ and $C_p$ for the NACA 0012 airfoil: a)in the whole computational domain; b) on airfoil surface}
\end{figure}

Figure~\ref{NACA_cp0} shows the distribution of the static pressure over the entire computational domain, while Fig.~\ref{NACA_cp} compares the surface pressure coefficient on the upper side of the airfoil obtained using the flow GKS solver with the experimental data. The numerical results exhibit good agreement with the measurements.

\subsection{Adjoint GKS Solver Verification}

\subsubsection{Durham Turbine Cascade Case}

To verify the adjoint GKS solver, the mass flow rate at the outlet is chosen as the objective function, and the inlet total temperature is selected as the design variable for this test case.

\begin{figure}[h!]
\centering
\subfigure[flow fields]{
	\includegraphics[width=2.0in]{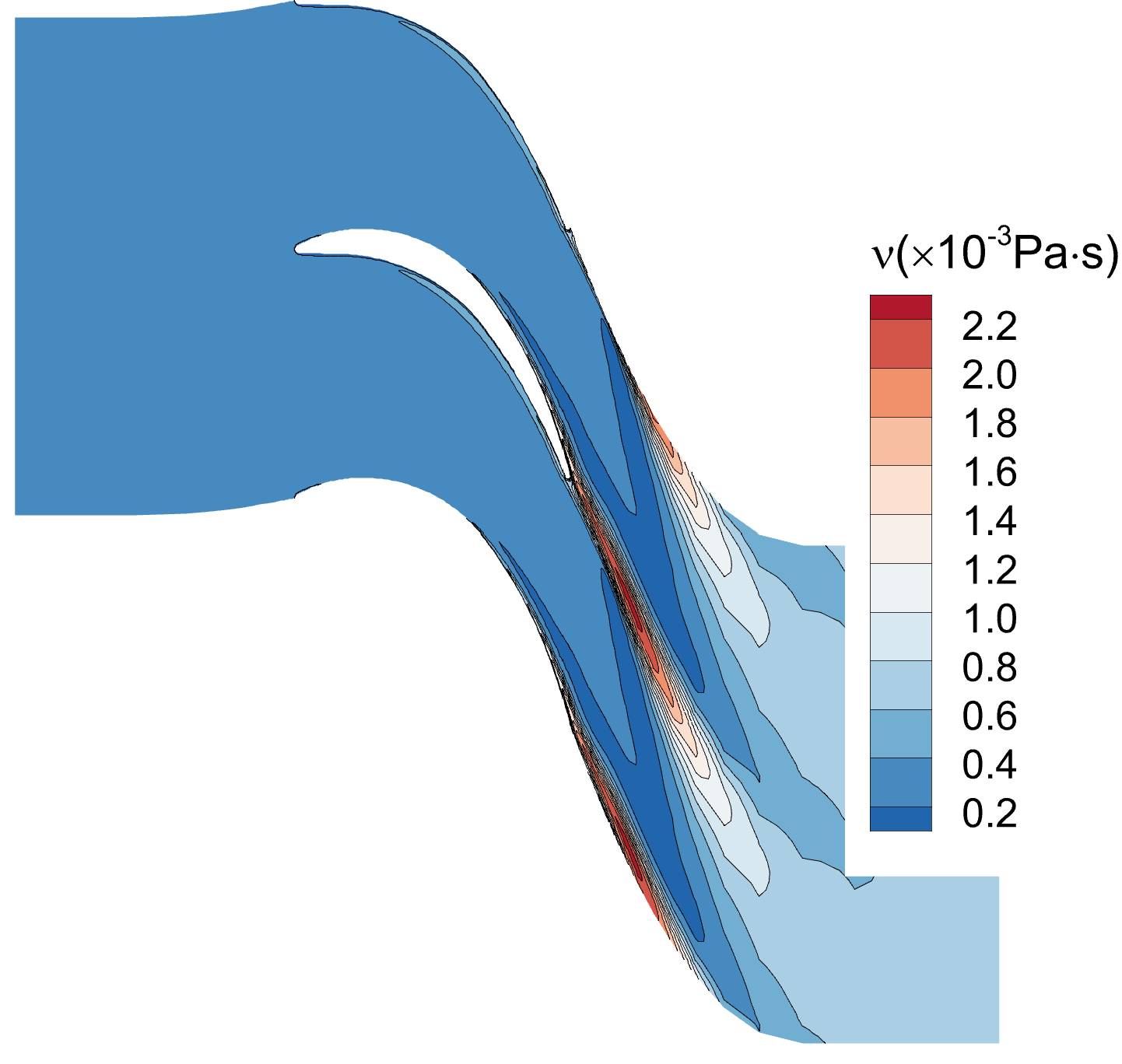}
 \label{durham_flow}
 }
 \subfigure[adjoint fields]{
	\includegraphics[width=2.0in]{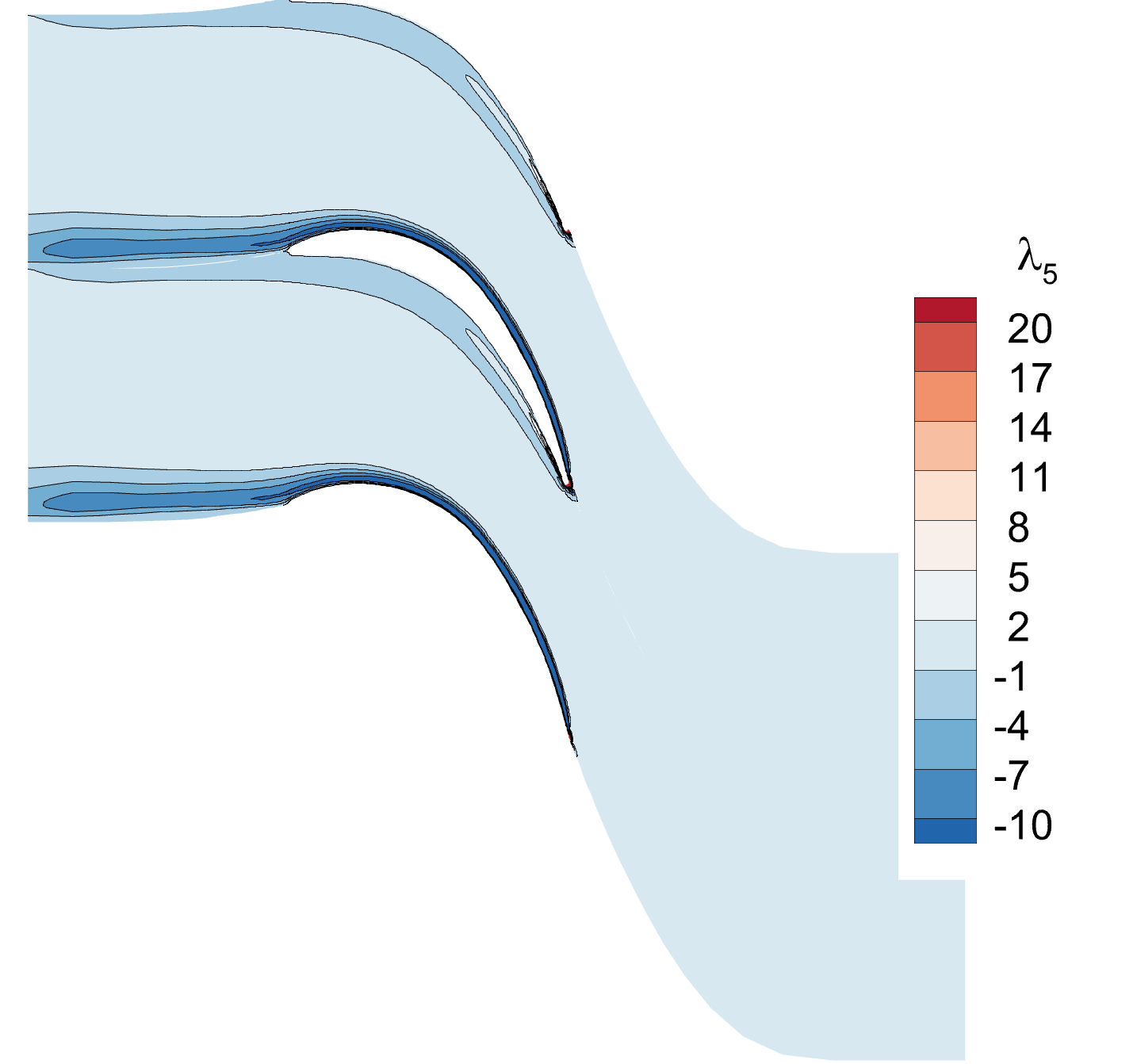}
 \label{durham_adj}
 }
	\caption{Both the flow and adjoint variable contours related to the turbulence model equation for the Durham turbine cascade case}
\end{figure}

A comparison of the flow and adjoint fields associated with the turbulence variable is presented in Figs.~\ref{durham_flow} and \ref{durham_adj}.
As expected, a distinct difference between the two fields is observed: the adjoint field exhibits an opposite trend relative to the flow field. For example, the turbulence variable behind the trailing edge is large in the flow solution, whereas the adjoint turbulence variable is large upstream of the leading edge. This behavior reflects the reverse mode of the adjoint principle and is commonly used as a qualitative verification of adjoint solvers.

\begin{figure}[h!]
\centering
\subfigure[sensitivity]{
	\includegraphics[width=2.0in]{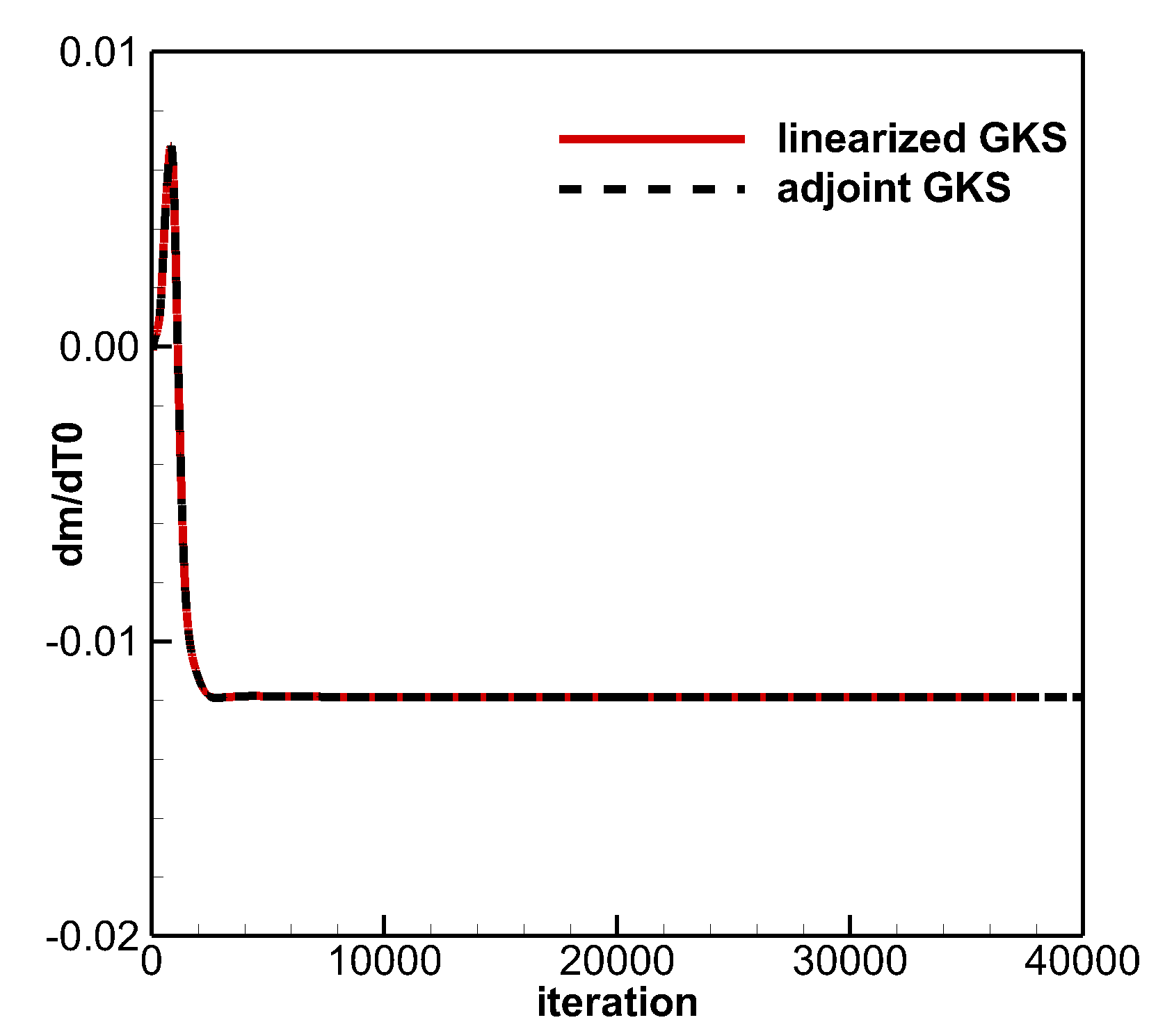}
 \label{durham_sen}
 }
 \subfigure[residual]{
	\includegraphics[width=2.0in]{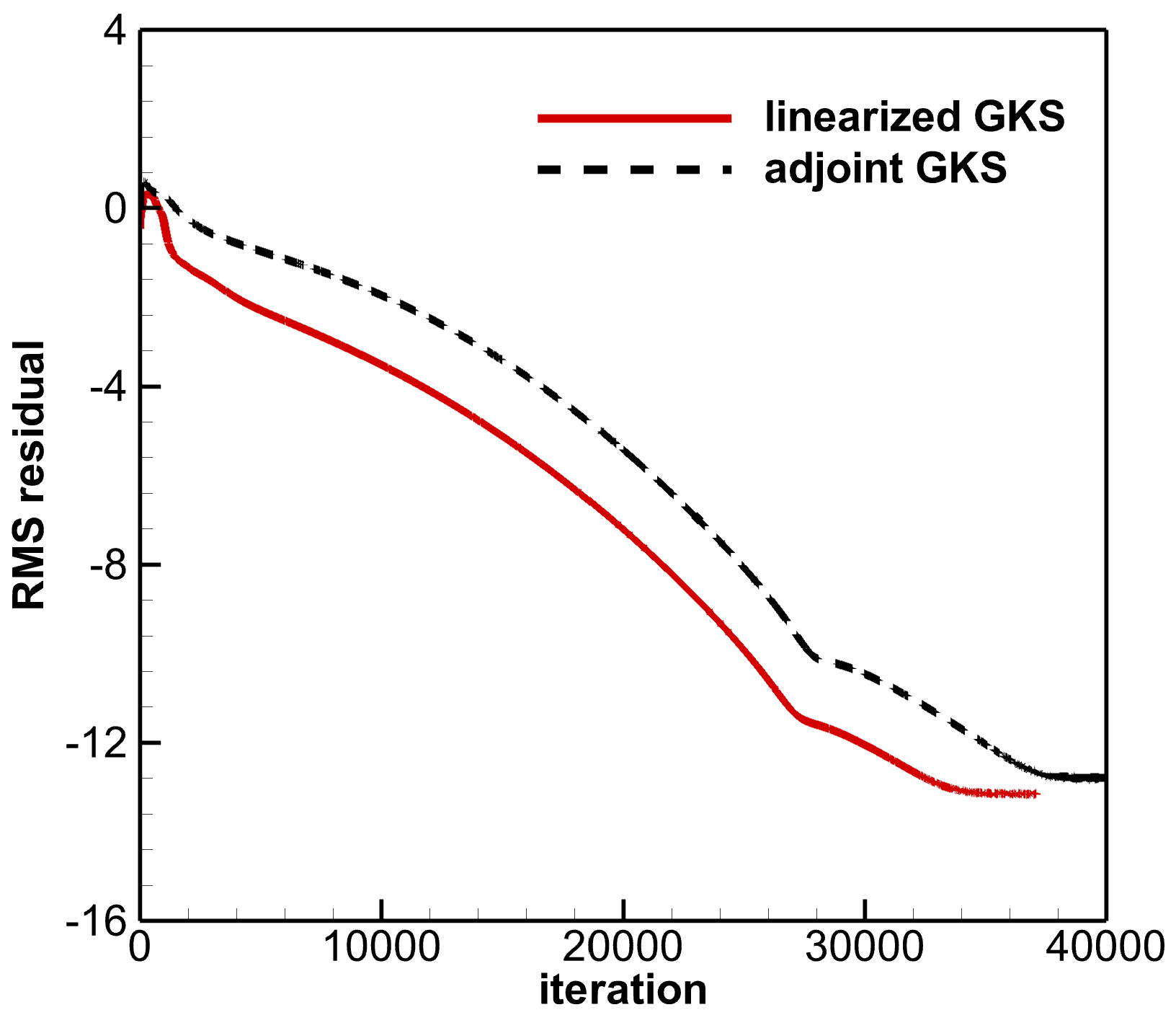}
 \label{durham_res}
 }
	\caption{Evolutionary histories of sensitivities and RMS residuals for both adjoint and linearized GKS solvers for the Durham turbine cascade case}
\end{figure}
Figures~\ref{durham_sen} and \ref{durham_res} compare the evolutionary histories of the sensitivities and RMS residuals for both the adjoint and linearized GKS solvers. The adjoint sensitivity convergence curve closely matches that of the linearized solver. At full convergence, the adjoint sensitivity is -0.0011895298, whereas the linearized sensitivity is -0.0011895300; the relative difference between the two values is negligible.
Furthermore, Figure~\ref{durham_res} shows that the two solvers exhibit the same slope in their RMS residual convergence curves, confirming their duality‑preserving behavior.

\subsubsection{NACA 0012 Airfoil}

\begin{figure}[h!]
\centering
\subfigure[flow fields]{
	\includegraphics[width=3.0in]{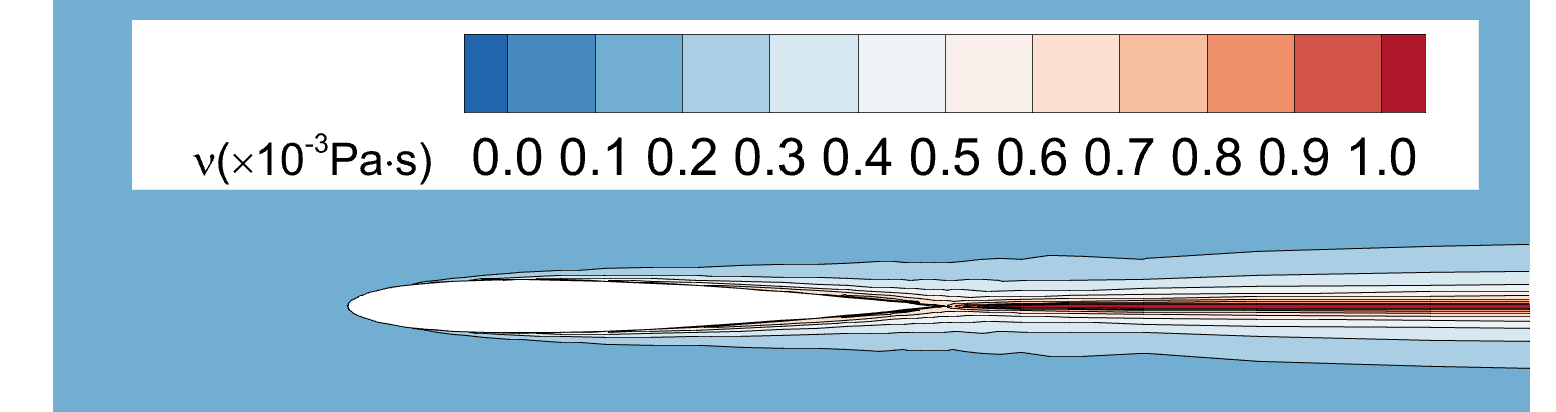}
 \label{naca_flow}
 }
 \subfigure[adjoint fields]{
	\includegraphics[width=3.0in]{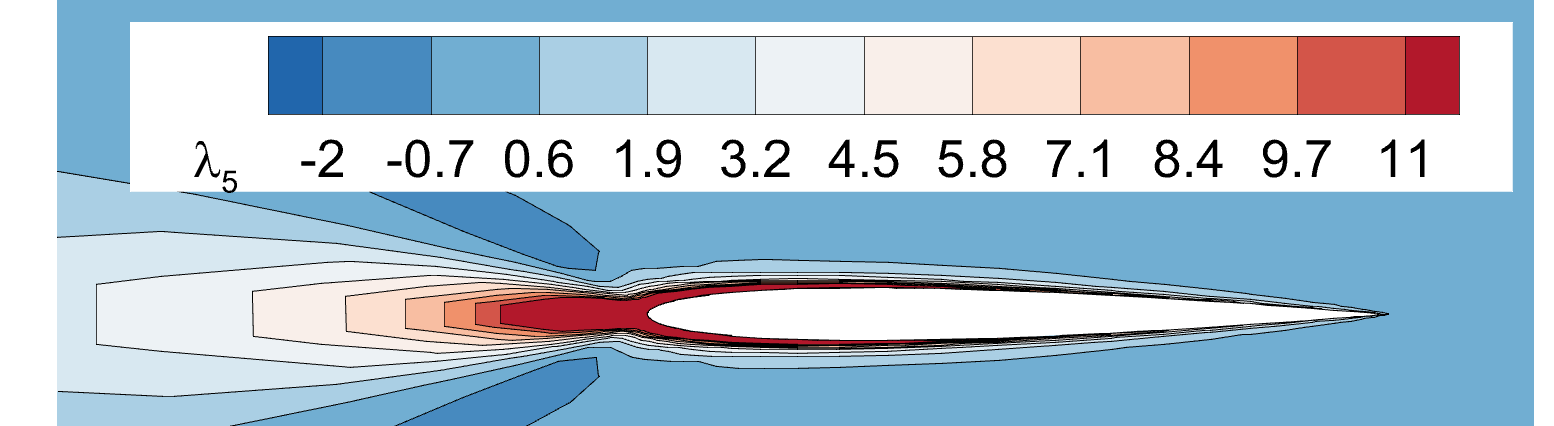}
 \label{naca_adj}
 }
	\caption{Both the flow and adjoint variable contours related to the turbulence model equation for the NACA 0012 airfoil}
\end{figure}
For this test case, the objective function is drag on the airfoil surface, and the design variable considered is the farfield static temperature. Figures~\ref{naca_flow} and \ref{naca_adj} show the turbulence variable contours and its adjoint counterpart. As expected, the turbulence variable is large downstream of the trailing edge, whereas the adjoint turbulence variable exhibits large values upstream of the leading edge. This opposite behavior between the flow and adjoint fields qualitatively demonstrates the correct implementation of the discrete adjoint GKS solver.

\begin{figure}[h!]
\centering
\subfigure[sensitivity]{
	\includegraphics[width=2.0in]{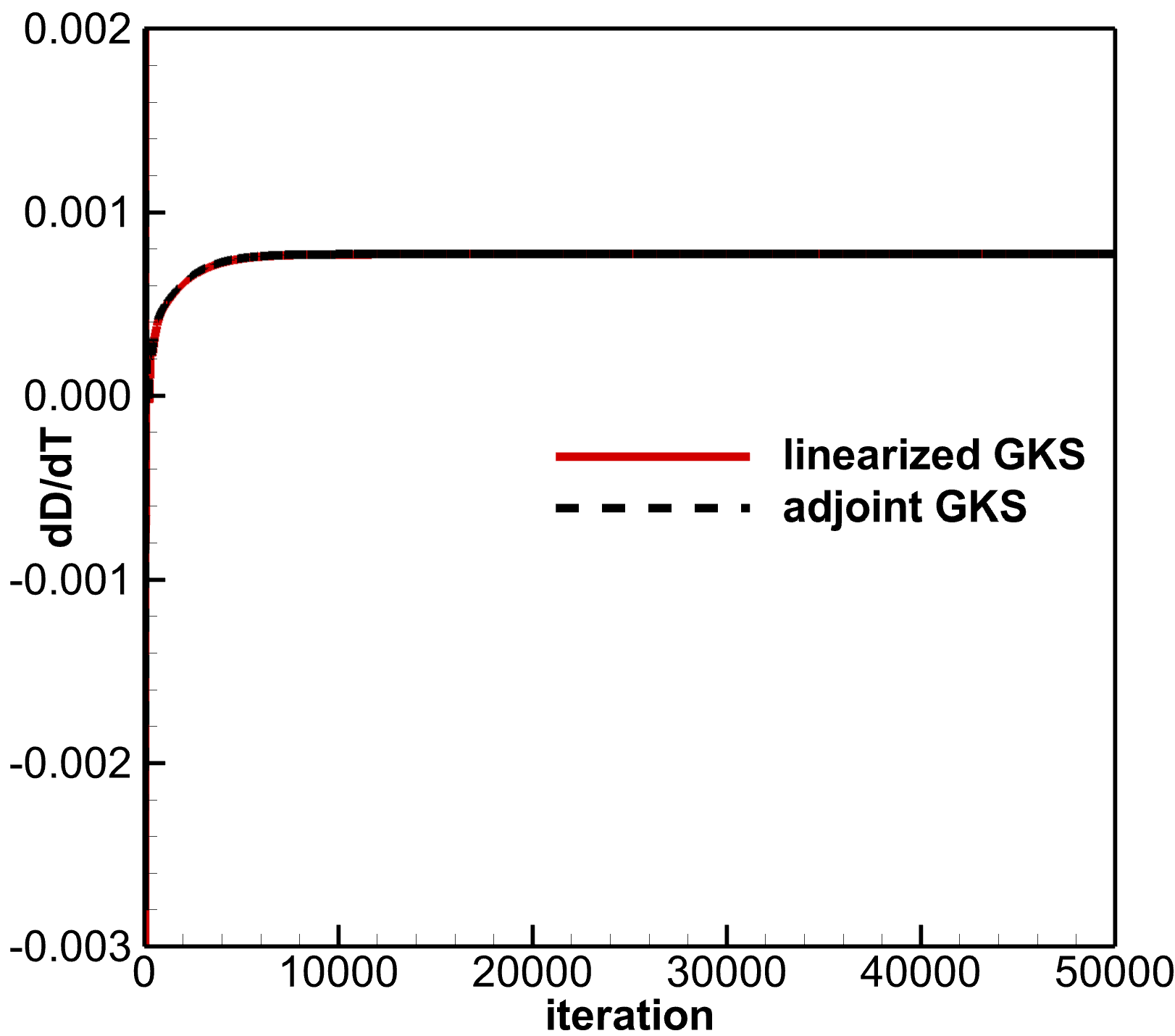}
 \label{naca_sen}
 }
 \subfigure[residual]{
	\includegraphics[width=2.0in]{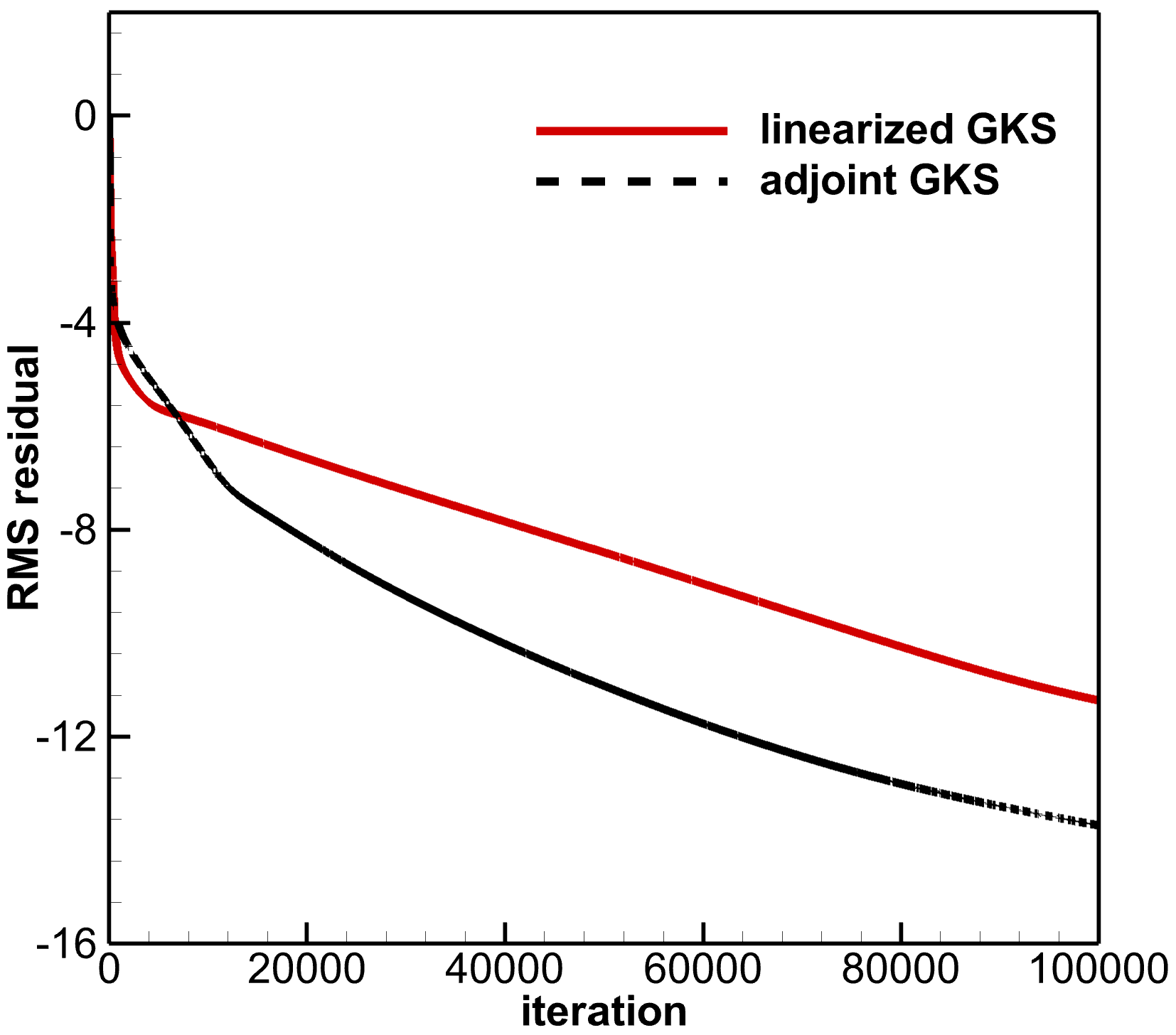}
 \label{naca_residual}
 }
	\caption{Evolutionary histories of sensitivities and RMS residuals for both adjoint and linearized GKS solvers for the NACA 0012 airfoil}
\end{figure}
To further quantitatively verify the adjoint solver, the convergence histories of both the sensitivities and the RMS residuals from the linearized and adjoint GKS solvers are compared in Figs.~\ref{naca_sen} and \ref{naca_residual}. The sensitivity convergence curve obtained using the adjoint solver overlaps with that of the linearized solver. In addition, the adjoint residual convergence curve is parallel to the linearized one. These results indicate that the adjoint GKS solver possesses the same eigenvalues as the linearized GKS solver. This duality‑preserving property provides quantitative confirmation of the correct development of the adjoint GKS solver.

\section{Results and Discussion}

In this section, the adjoint-based design optimization system and the functionality of each part are described in detail, followed by a discussion of the optimization results.

\subsection{Adjoint-Based Design Optimization System}

\begin{figure}[h!]
    \centering
\includegraphics[width=0.7\linewidth]{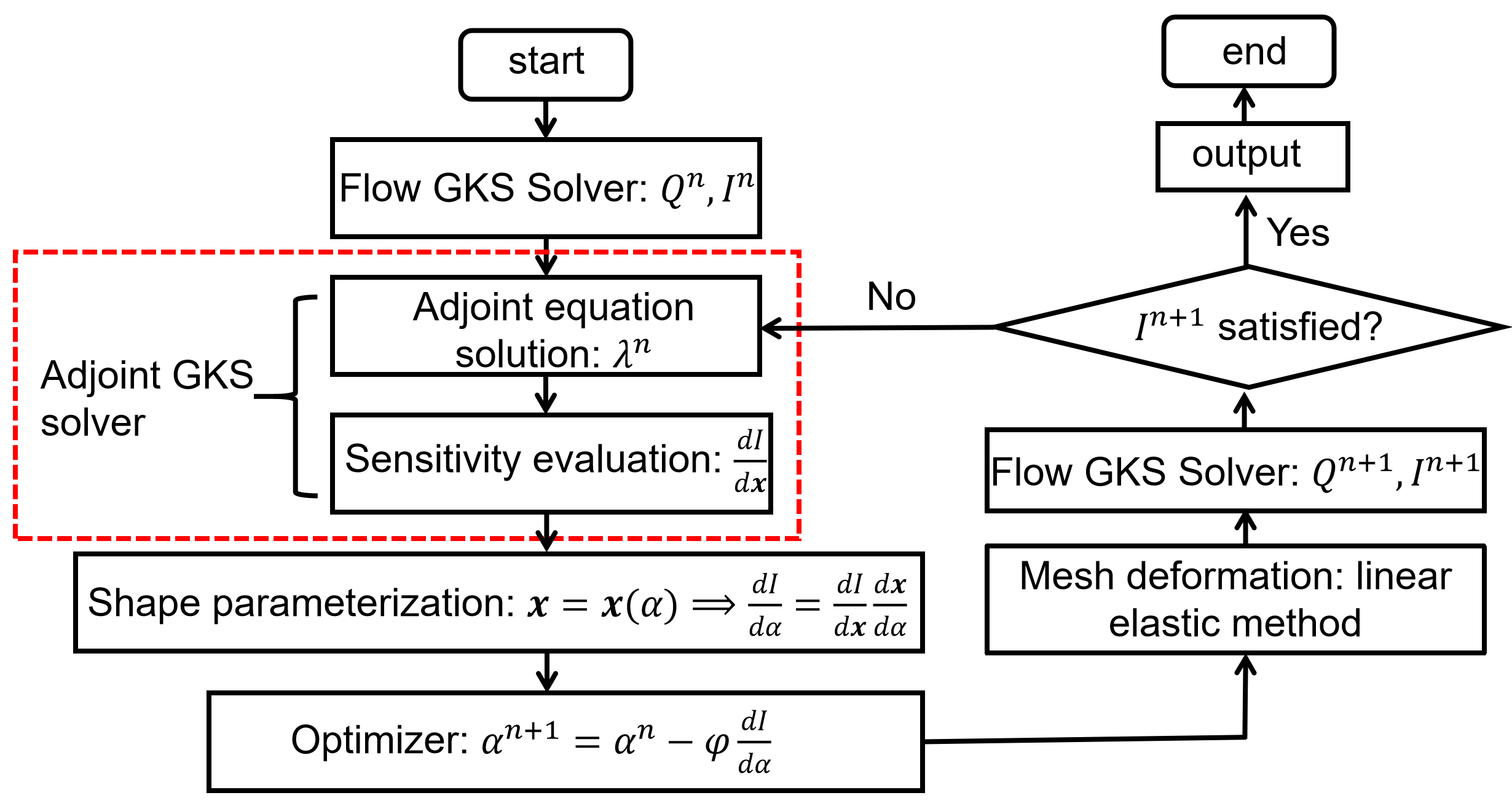}
    \caption{The dataflow of the adjoint-based design optimization system}
    \label{adj_opt}
\end{figure}

Figure \ref{adj_opt} shows the dataflow of the adjoint-based design optimization system. It contains five main parts: flow GKS solver, adjoint GKS solver, optimizer, shape parameterization, and mesh deformation.
\begin{enumerate}
\item flow GKS solver: compute the macroscopic flow variables and the objective function;

\item adjoint GKS solver: include adjoint equation solution and sensitivity evaluation. The former computes the adjoint variables, which are then supplied to the sensitivity‑evaluation module to obtain $\frac{dI}{d\boldsymbol x}$.

\item shape parameterization: construct the relationship between the grid coordinates ($\boldsymbol x$) and the design variables ($\alpha$). In this work, the Hicks-Henne hump function is used to parameterize the shape perturbation. It can be expressed by
\begin{equation}
\delta = \sum_{i=1}^{N}\alpha_i sin^4(\pi x_d^{n_i})(i=1,2,...,N)
\end{equation}
where $N$ is the number of design variables, $x_d$ and $n_i$ are defined by
\begin{equation}
x_d = \frac{x-x_t}{x_e-x_t}, n_i = \frac{ln0.5}{ln m_i},m_i=\frac{i}{N+1}
\end{equation}
Figure \ref{Hicks-Henne} presents the distribution of five sets of Hicks-Henne hump functions. It can be seen that the Hicks-Henne hump function has very obvious localization property.
\begin{figure}[h!]
    \centering
\includegraphics[width=0.4\linewidth]{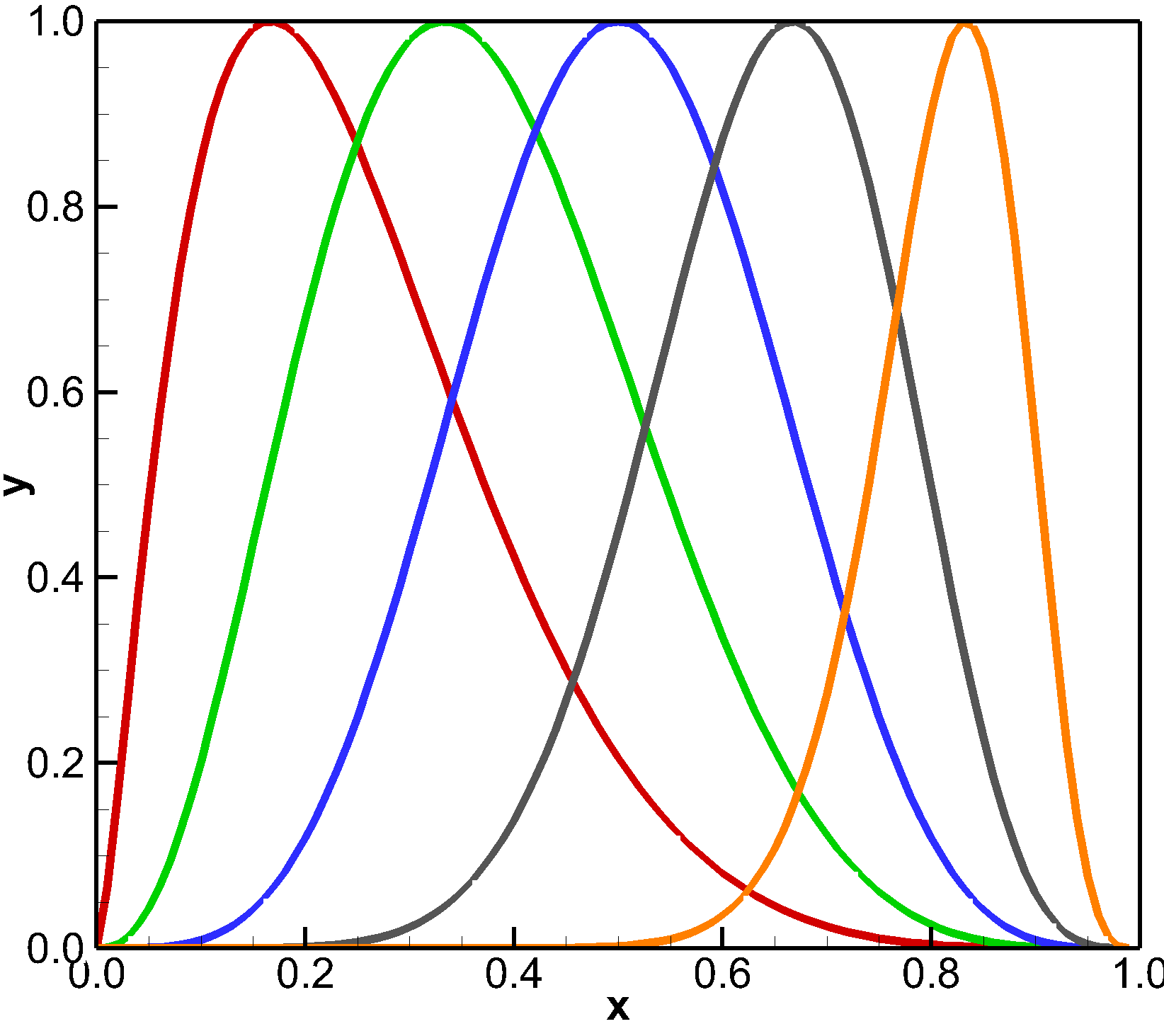}
    \caption{Five sets of Hicks-Henne hump functions}
    \label{Hicks-Henne}
\end{figure}

Based on the chain rule, the sensitivity of $I$ with respect to $\alpha$ can be obtained, as shown in the following equation
\begin{equation}
\frac{dI}{d\alpha} = \frac{dI}{d\boldsymbol x}\frac{d\boldsymbol x}{d\alpha}
\end{equation}

\item optimizer: update the design variables.
\begin{equation}
\alpha^{n+1} = \alpha^{n} - \psi \frac{dI}{d\alpha}
\end{equation}
where $\psi$ represents the step size, determined by the optimization method. In this work, the steepest descent method with a constant step size is used.

\item mesh deformation: regenerate mesh for the optimized geometry. To guarantee mesh quality, the linear elastic method is applied in this work.
\begin{equation}
\Delta y_{new} = (1-r)\Delta y_s + r\Delta y_f
\end{equation}
where $\Delta y_f$ represents the displacement of the coordinate at the far field boundary and is set to zero in this work, $\Delta y_s$ represents the displacement of the coordinate on the solid surface and is determined by optimization, $r$ is a weighting factor, defined by
\begin{equation*}
r = \frac{y - y_s}{y_f - y_s}
\end{equation*}
Figure \ref{mesh_deformation} compares the meshes between the original and optimized ones. It can be seen that the mesh quality is well preserved.

\begin{figure}[h!]
    \centering
\includegraphics[width=0.6\linewidth]{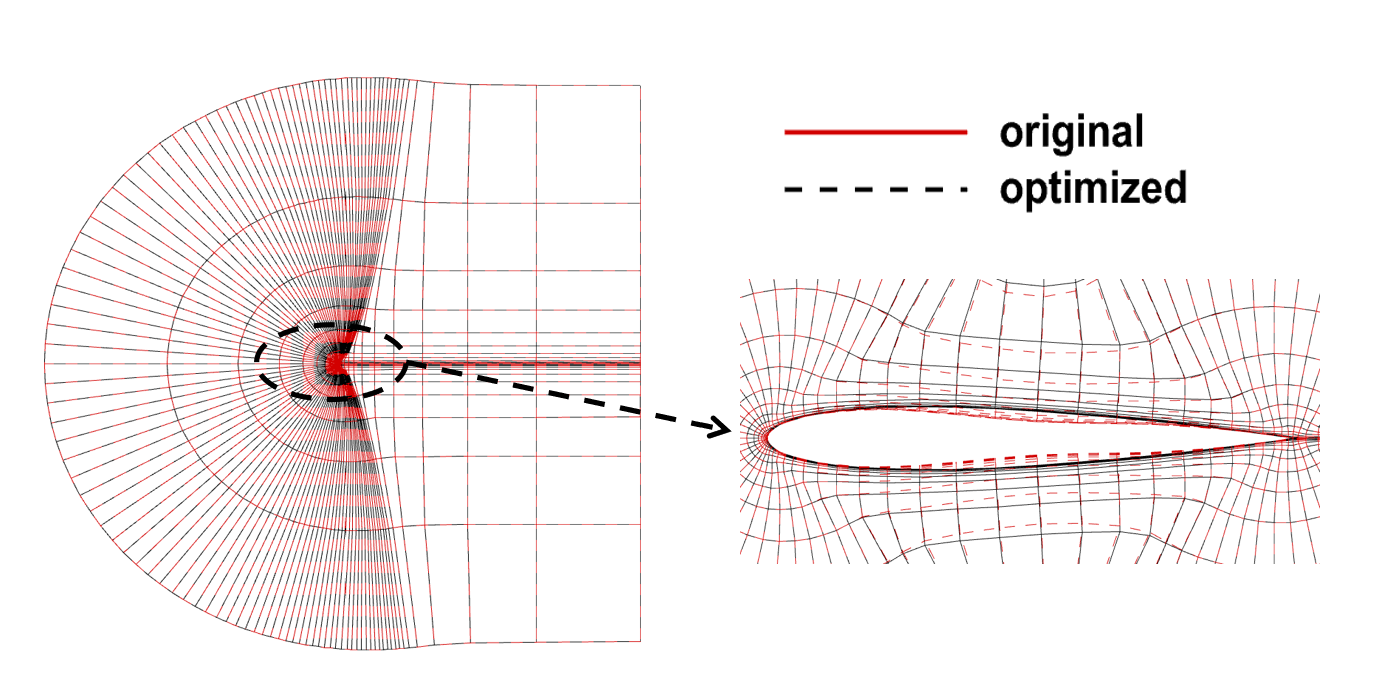}
    \caption{Comparison of mesh quality between the original and optimized airfoils.}
    \label{mesh_deformation}
\end{figure}

\end{enumerate}

Perform the flow analysis again to compute the updated flow field information and objective function. If the convergence criteria are satisfied, the optimized geometry is obtained. Otherwise, the above procedure is repeated until the required convergence is achieved.

\subsection{Design Optimization}

\subsubsection{Inverse Design of Turbine Blades}

To demonstrate the effectiveness of the adjoint GKS‑based design optimization system, the first test case is an inverse design problem. The original Durham turbine blade is first perturbed, leading to a Mach number distribution on the blade surface that differs from that of the original configuration. The objective of the optimization is to recover the original blade profile from the perturbed one based on the Mach number discrepancy, using the adjoint GKS optimization method. Accordingly, the objective function is defined as
\begin{equation}
I = \sum_{i=1}^{M}(Ma_i-Ma_{i,0})^2
\end{equation}
where $I$ is the objective function, $M$ is the number of grid points on the blade surface; $Ma$ is the isentropic Mach number computed by the static pressure on the blade surface and the inlet total pressure and the subscript 0 represents the target Mach number distribution.

The design variables are the coefficients of the Hicks–Henne hump functions, which are used to parameterize the perturbation of the camber. In this case, eight design variables are employed and are uniformly distributed on the camber.

\begin{figure}[h!]
    \centering
\includegraphics[width=0.4\linewidth]{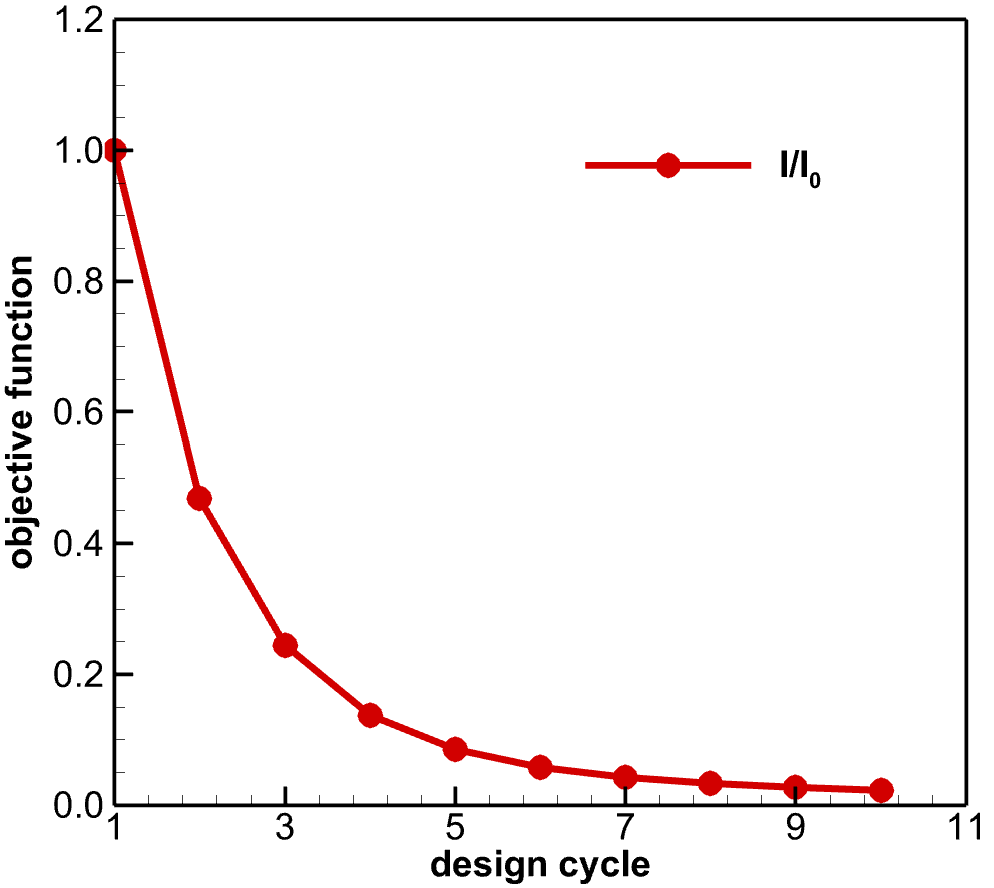}
    \caption{Evolutionary history of the objective function for the Durham turbine cascade case}
    \label{durham_obj}
\end{figure}
\begin{figure}[h!]
\centering
\subfigure[Mach number]{
	\includegraphics[width=2.0in]{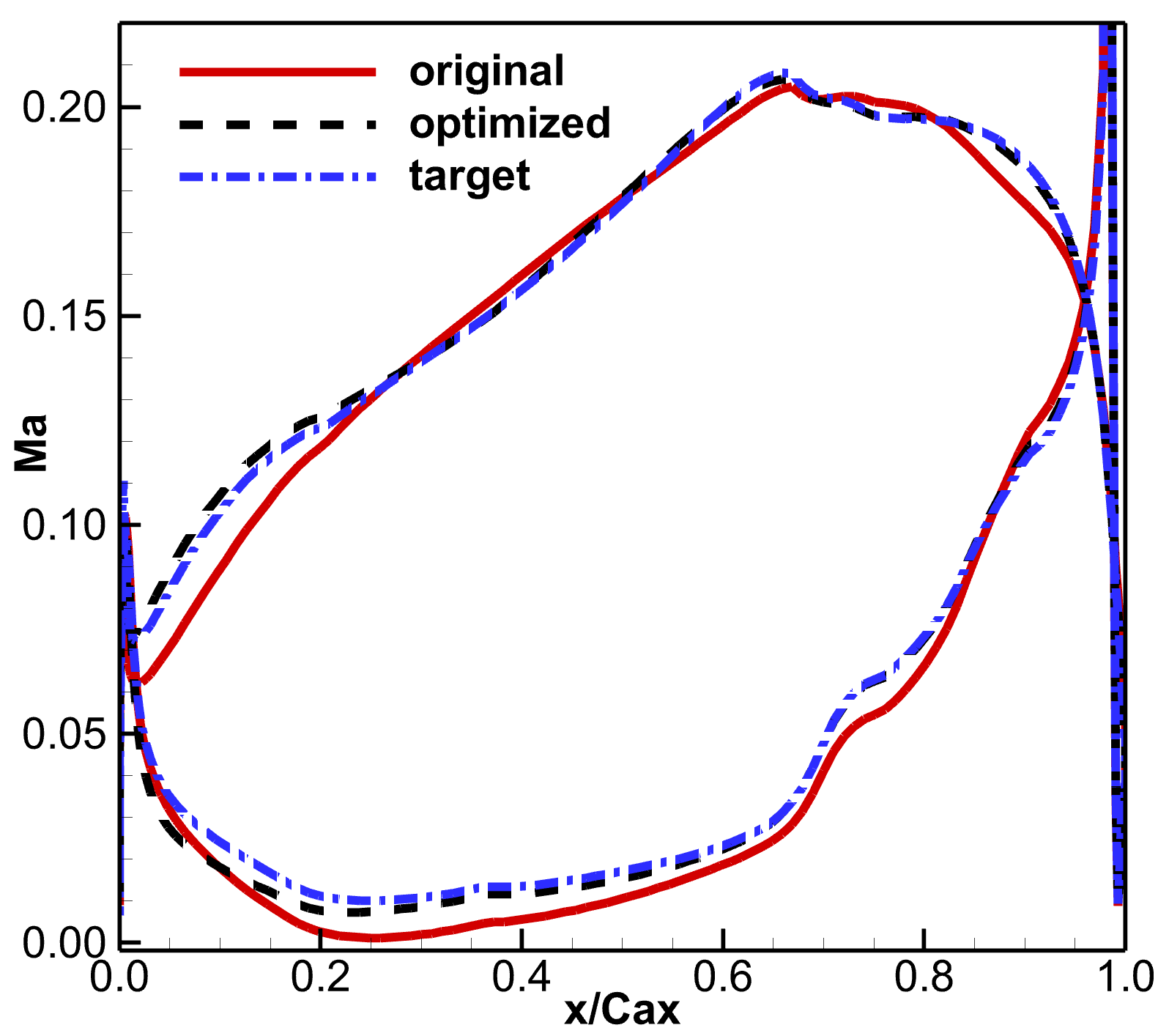}
 \label{durham_ma_opt}
 }
 \subfigure[blade profile]{
	\includegraphics[width=2.0in]{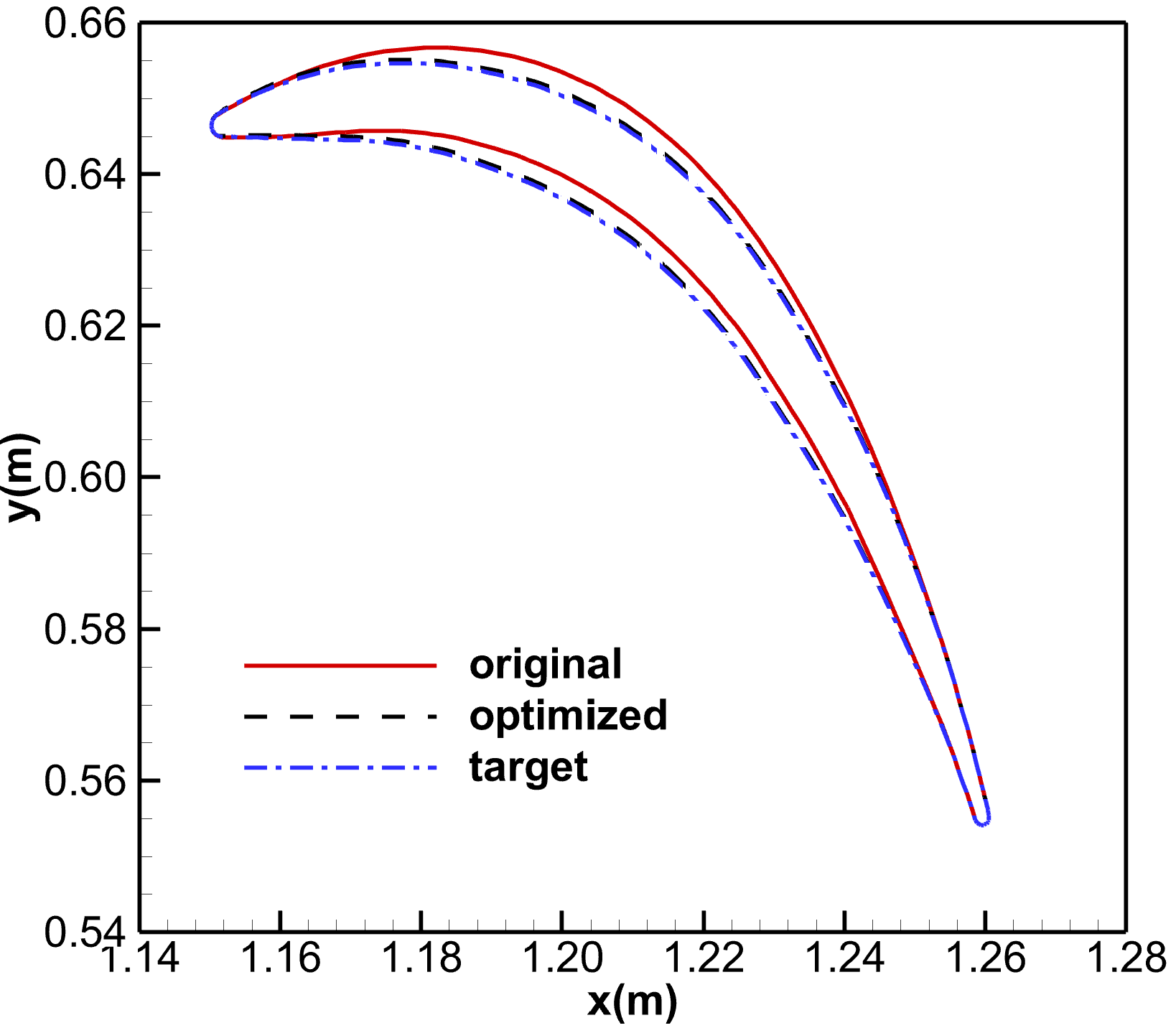}
 \label{durham_blade}
 }
	\caption{Comparison of the Mach number distributions on blade surfaces and blade profiles among the original, optimized, and target ones for the Durham turbine cascade case}
\end{figure}

Figure~\ref{durham_obj} shows the evolutionary history of the objective function. After ten design cycles, the objective function is reduced by 99.9\%, indicating that the Mach number discrepancy between the optimized and target distributions has become negligible. The distributions of the Mach number on the blade surface, as shown in Fig.~\ref{durham_ma_opt}, further demonstrate this point. Figure~\ref{durham_blade} presents a comparison among the original, optimized, and target blade profiles. The results show that the inverse design procedure using the adjoint GKS-based optimization method accurately reconstructs the target profile from the perturbed geometry (denoted as ‘original’ in the legend).

\subsubsection{Lift-to-Drag Ratio Enhancement}

The lift-to-drag ratio is commonly used as a quantitative measure of an airfoil’s aerodynamic performance, with higher values indicating better performance. Accordingly, the lift-to-drag ratio is adopted as the objective function in the present optimization.
\begin{equation}
I = \hat L/\hat D
\end{equation}
Where $\hat L(=L/L_0)$ is the non-dimensional lift  and $\hat D(=D/D_0)$ represents the non-dimensional drag.

Since the original NACA 0012 airfoil is symmetric, its lift is zero at a zero angle of attack. In this optimization, the angle of attack is set to $1^{\circ}$, while all other parameters are kept identical to those used in the solver-validation section. Furthermore, for this design optimization, ten design variables are used including five uniformly distributed on the upper surface and five on the lower surface.

\begin{figure}[h!]
    \centering
\includegraphics[width=0.4\linewidth]{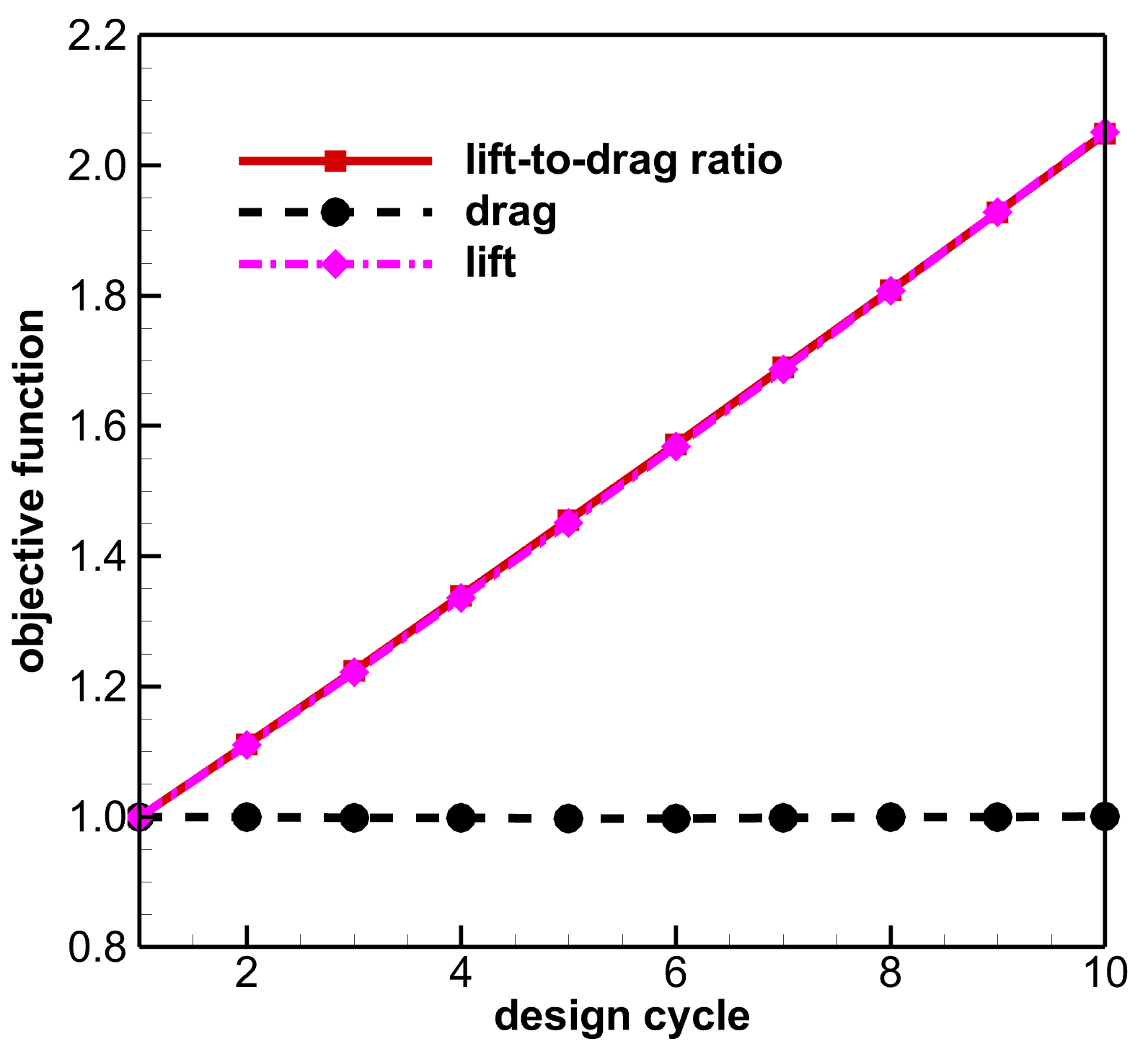}
    \caption{Evolutionary history of the objective function for the NACA 0012 airfoil}
    \label{naca_obj1}
\end{figure}
\begin{figure}[h!]
    \centering
\includegraphics[width=0.4\linewidth]{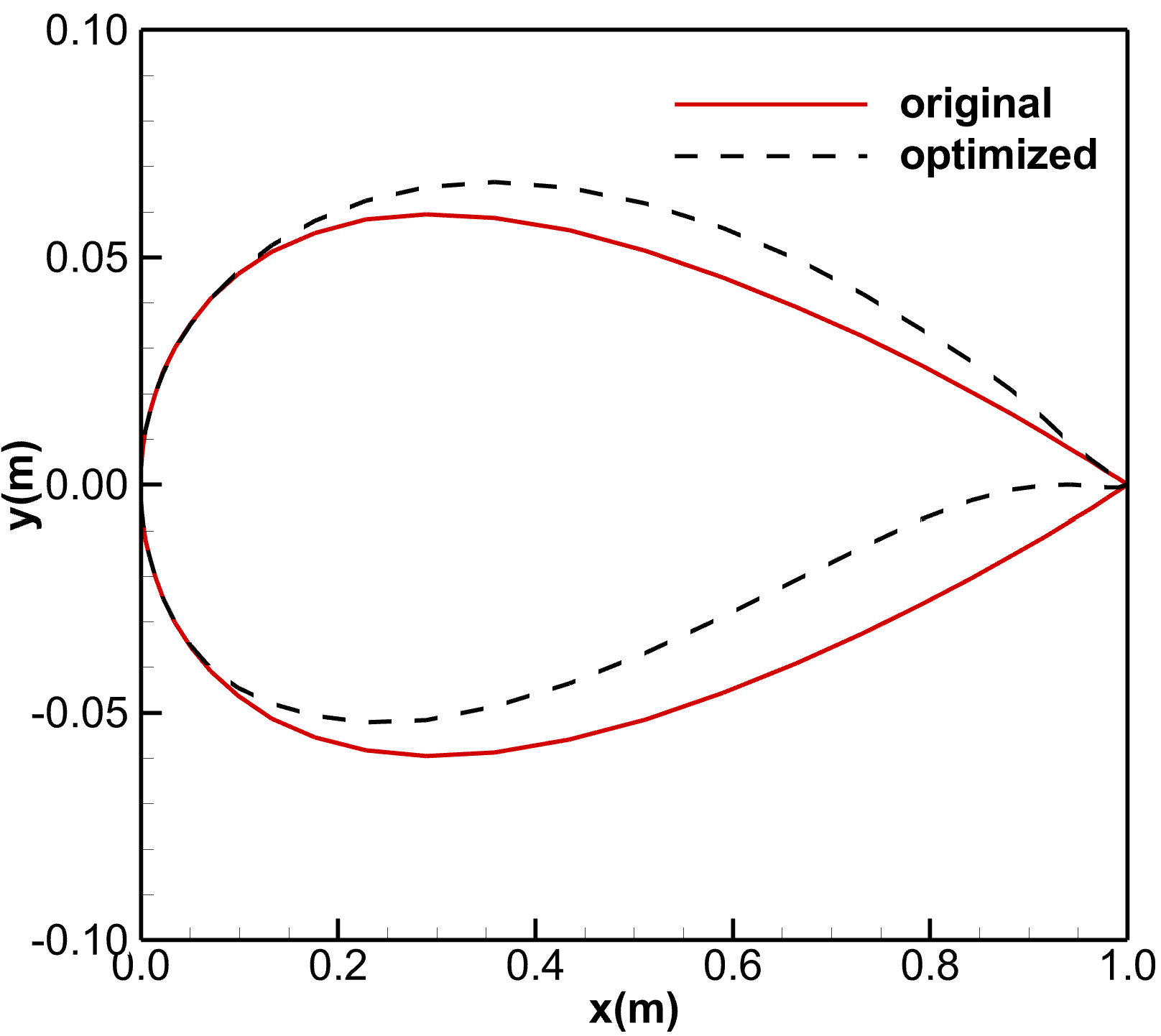}
    \caption{Comparison of original and optimized NACA 0012 airfoils}
    \label{naca_airfoil1}
\end{figure}

Figure~\ref{naca_obj1} presents the evolutionary histories of the objective functions. Compared with the original configuration, the drag remains essentially unchanged, while the lift increases by more than a factor of two after ten design cycles. As a result, the lift‑to‑drag ratio also shows a noticeable upward trend. Figure~\ref{naca_airfoil1} compares the original and optimized NACA 0012 airfoils. Because the design variables on the upper and lower surfaces influence the lift‑to‑drag ratio differently, the optimized airfoil becomes asymmetric, particularly near the trailing edge.

\begin{figure}[h!]
    \centering
\includegraphics[width=0.6\linewidth]{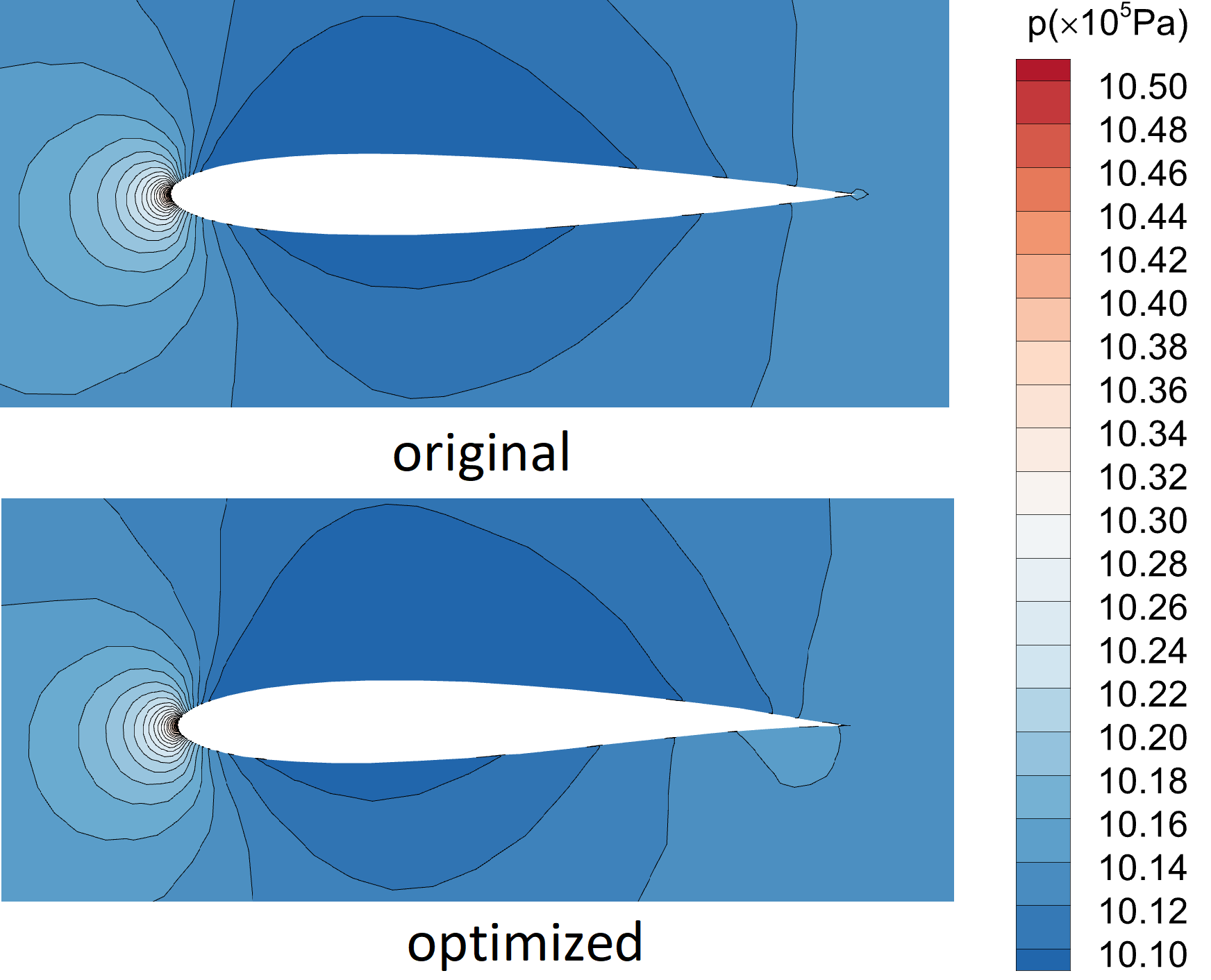}
    \caption{Comparison of the static pressure contours between the original and optimized NACA 0012 airfoils}
    \label{naca_opt_flow_field1}
\end{figure}

\begin{figure}[h!]
    \centering
\includegraphics[width=0.4\linewidth]{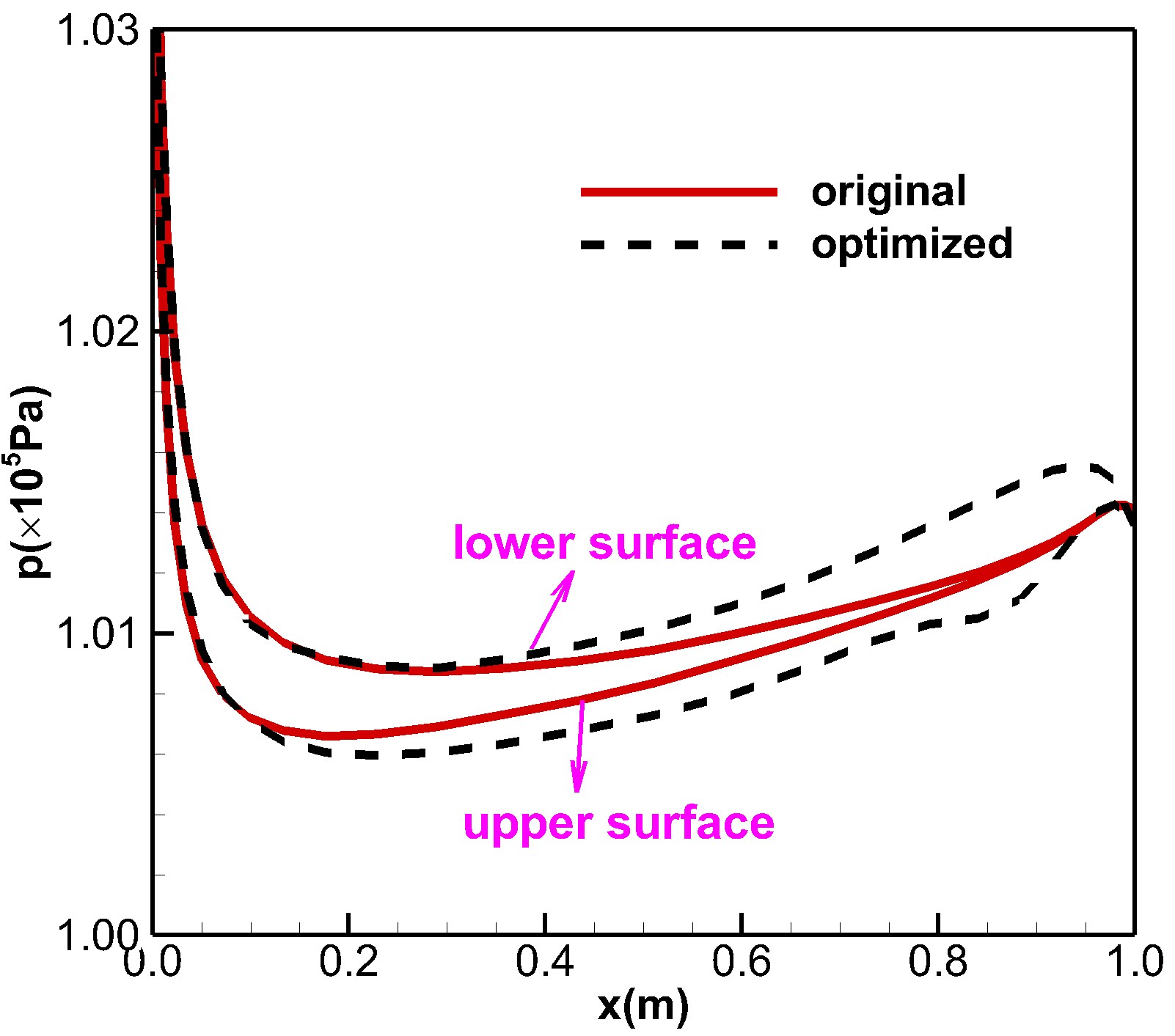}
    \caption{Comparison of the static pressure on surface between the original and optimized NACA 0012 airfoils}
    \label{naca_opt_p1}
\end{figure}

Figure~\ref{naca_opt_flow_field1} compares the static pressure contours in the entire computational domain for the original and optimized NACA 0012 airfoils. A clear difference can be observed: after optimization, the static pressure decreases on the upper surface and increases significantly on the lower surface, especially near the trailing edge. This trend is further confirmed in Fig.~\ref{naca_opt_p1}, which compares the surface pressure distributions. A reduction in static pressure on the upper surface combined with an increase on the lower surface is favorable for enhancing lift.

\subsubsection{Shock-Strength Reduction}

In this optimization, the NACA 0012 airfoil is considered again. The farfield Mach number is set to 0.8 and the angle of attack to zero, while the remaining parameters are identical to those in the solver‑validation section. Under these conditions, a shock wave forms within the flow field. The purpose of the optimization is to minimize the strength of this shock.
Since shock waves induce significant entropy generation—and stronger shocks correspond to higher entropy—the entropy‑generation rate may be used as an objective function for reducing shock intensity. However, designing a suitable objective function for this purpose requires careful consideration.

\begin{figure}[h!]
\centering
\subfigure[viscous]{
	\includegraphics[width=2.5in]{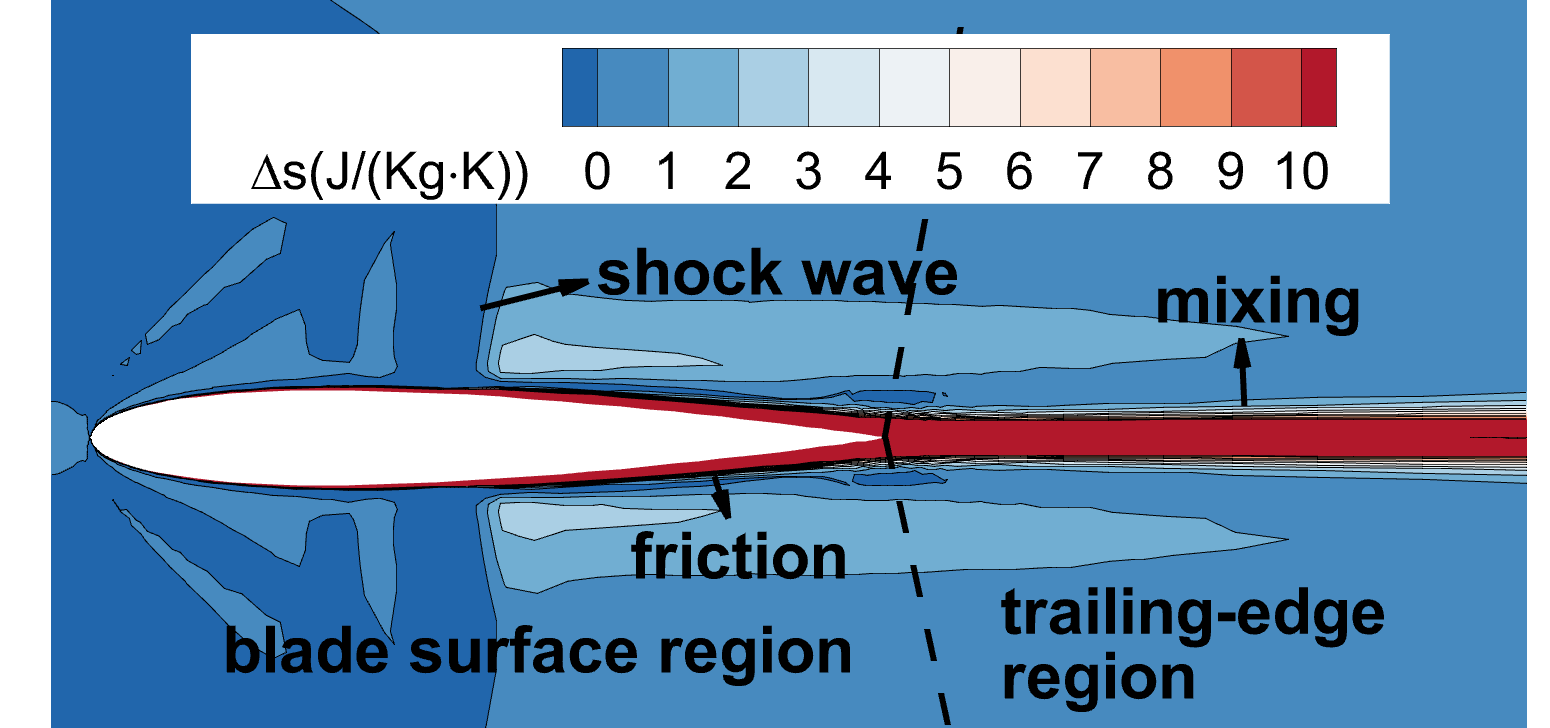}
 \label{naca_ds_sa}
 }
 \subfigure[inviscid]{
	\includegraphics[width=2.5in]{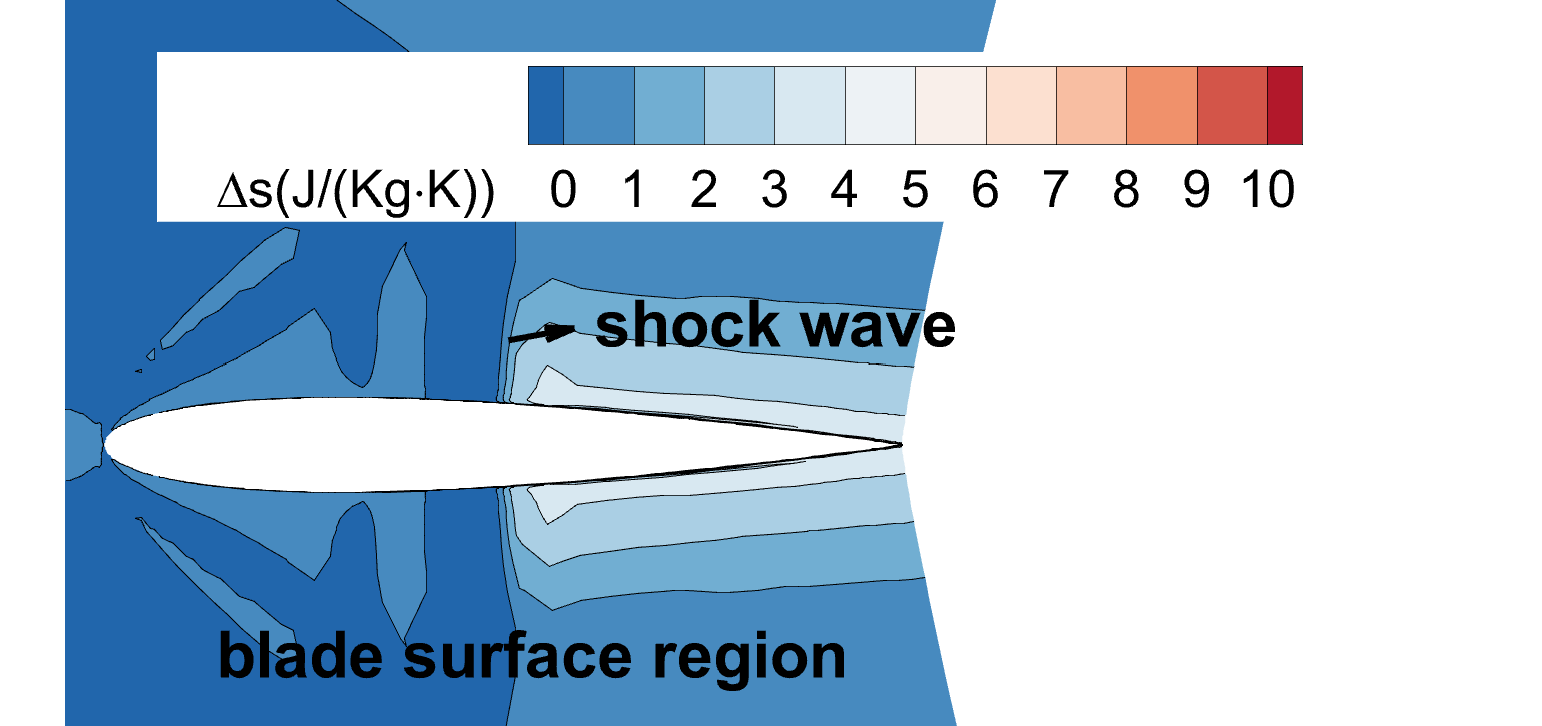}
 \label{naca_ds_euler}
 }
	\caption{Entropy contours for the NACA 0012 airfoil: a) viscous flow; b) inviscid flow}
\end{figure}

Figure~\ref{naca_ds_sa} presents the entropy contours of the viscous flow across the entire computational domain. The results indicate that entropy generation primarily arises from three mechanisms: the shock wave, the mixing occurring downstream of the trailing edge, and the friction on the solid wall. These contributions can be summarized by the following equation.
\begin{equation}
\Delta s = \sum\Delta s_{friction} + \sum\Delta s_{mixing} +\sum \Delta s_{shock}
\end{equation}

To minimize the entropy generated by the shock wave, it is necessary to carefully design the objective function. First, to eliminate the entropy contribution from wall friction, the optimization will employ the Euler equations together with a slip-wall boundary condition. Second, to suppress the entropy generated by flow mixing, only the entropy within the blade surface region will be considered in the objective function. The entropy distribution used in the evaluation of the objective function is shown in Fig.~\ref{naca_ds_euler}. The expression of the objective function is given by
\begin{equation}
I =\sum \Delta s_{shock}
\end{equation}

In this optimization, five design variables are uniformly distributed on the upper surface of the airfoil. The shape modifications applied to the lower surface are taken to be the opposite of those on the upper surface. As a result, the optimized airfoil is expected to retain its symmetric profile.

\begin{figure}[h!]
    \centering
\includegraphics[width=0.4\linewidth]{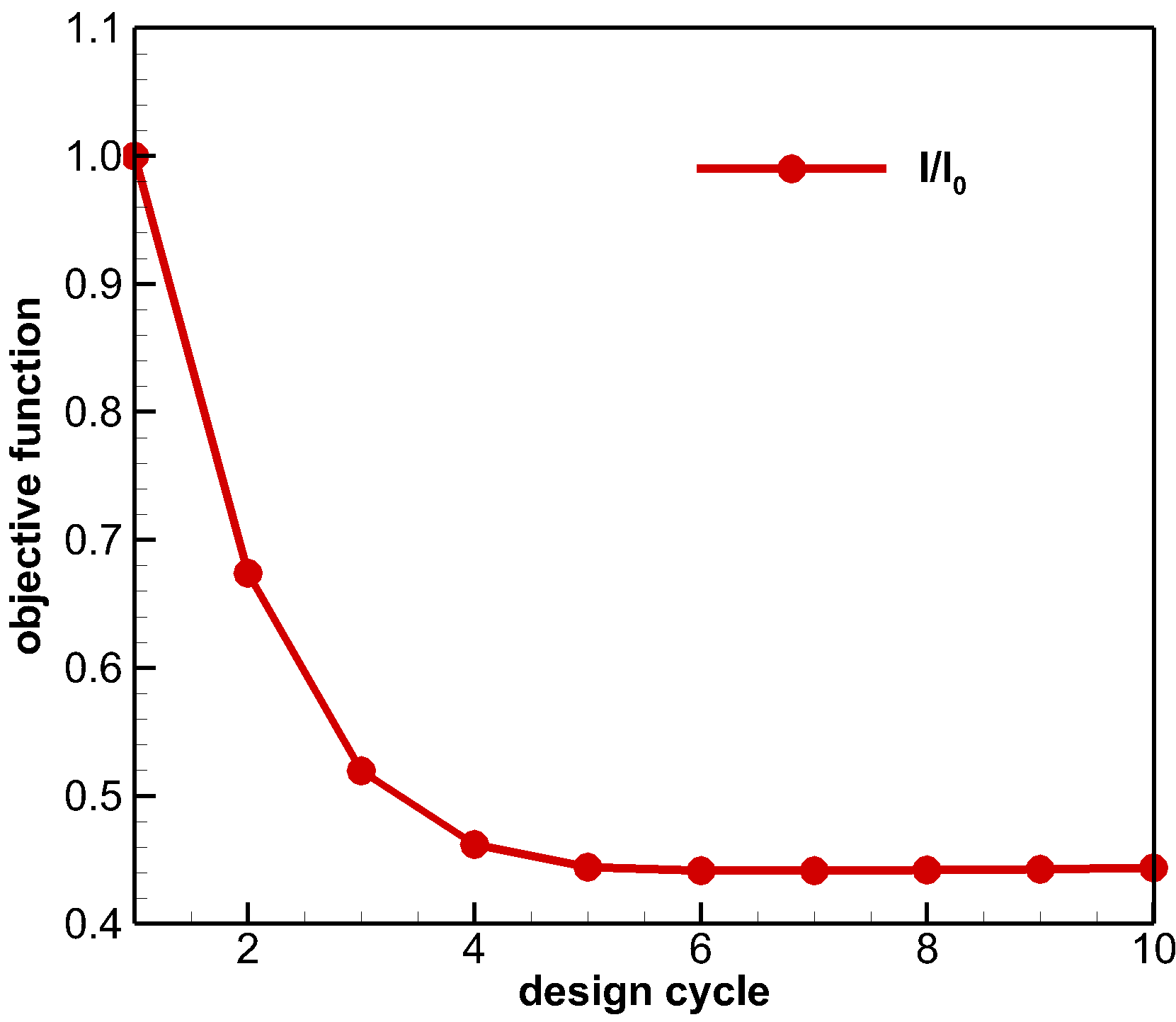}
    \caption{Evolutionary history of the objective function for reducing shock strength of the NACA 0012 airfoil}
    \label{naca_obj2}
\end{figure}
\begin{figure}[h!]
    \centering
\includegraphics[width=2.5in]{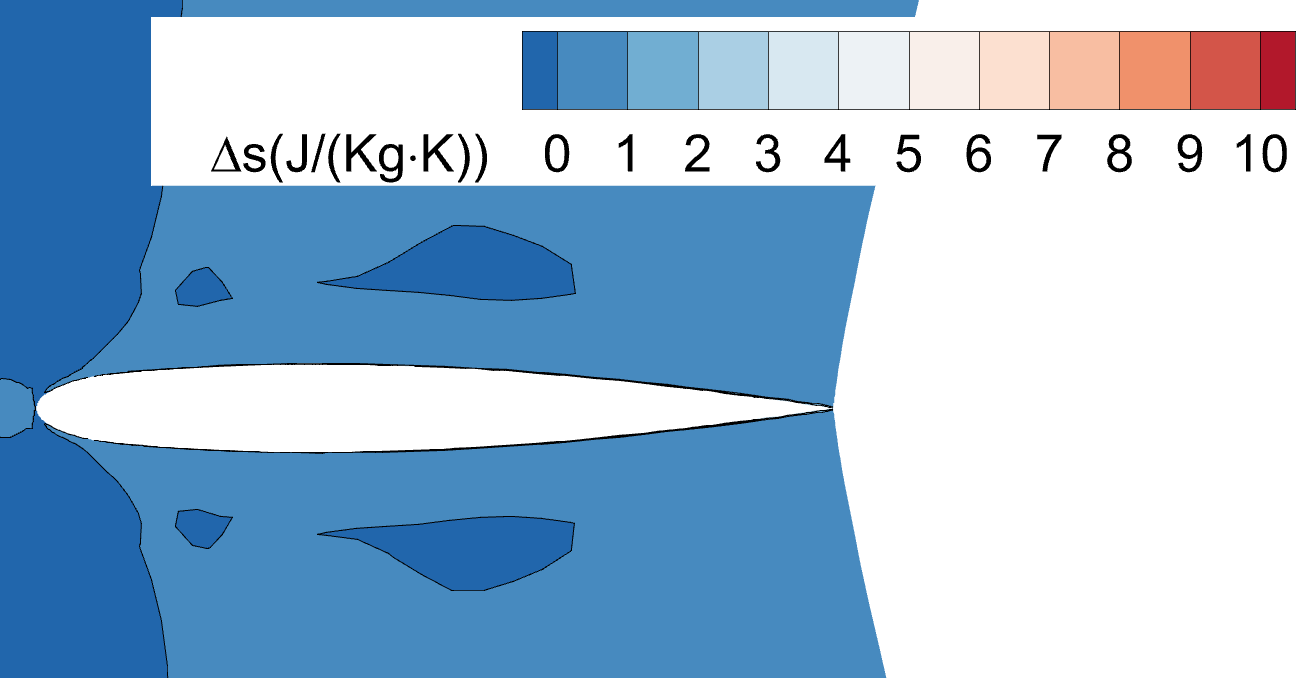}
    \caption{The entropy contours for the optimized NACA 0012 airfoil}
    \label{naca_ds_euler_opt}
\end{figure}

Figure~\ref{naca_obj2} shows the evolutionary history of the objective function during the shock-strength–reduction optimization of the NACA 0012 airfoil. After ten design cycles, the entropy is reduced by more than 55\%. The entropy contours for the optimized airfoil are presented in Fig.~\ref{naca_ds_euler_opt}. Compared with the original entropy contours, as shown in Fig.~\ref{naca_ds_euler}, the optimized design exhibits a significant decrease in the entropy generation associated with the shock wave.

\begin{figure}[h!]
\centering
\subfigure[original]{
	\includegraphics[width=2.0in]{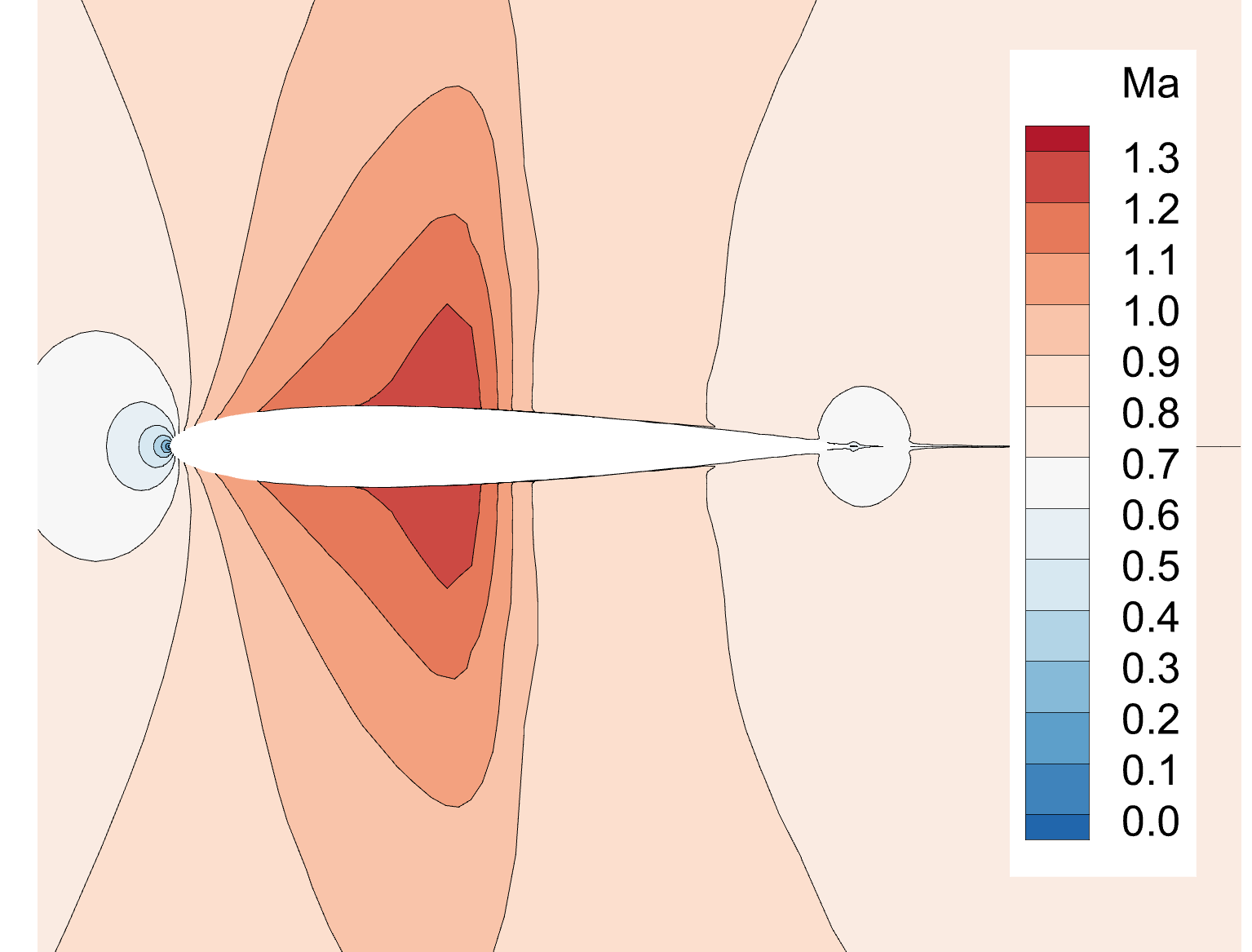}
 \label{naca_Ma_org}
 }
 \subfigure[optimized]{
	\includegraphics[width=2.0in]{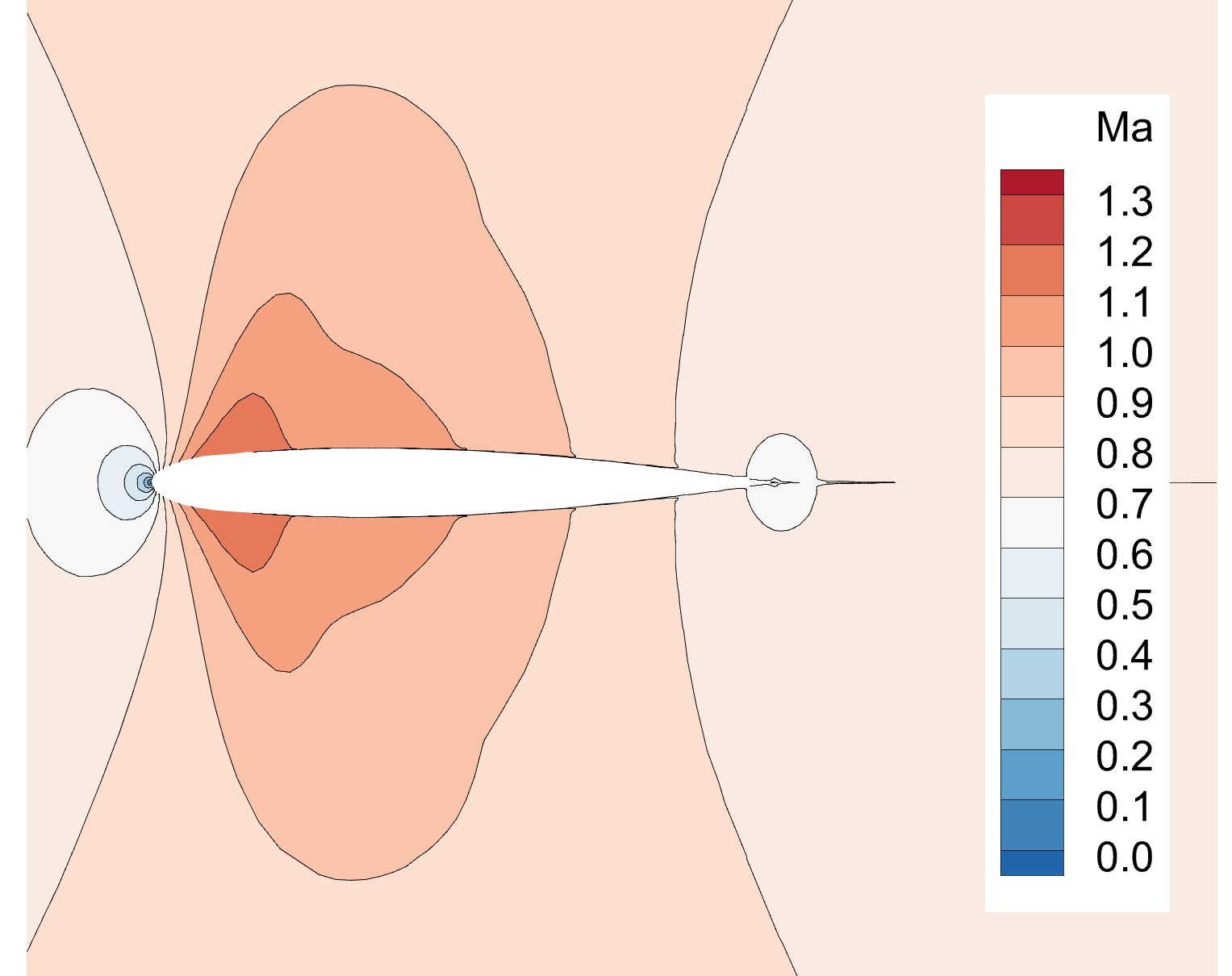}
 \label{naca_Ma_opt}
 }
	\caption{Comparison of the Mach number contours between the original and optimized NACA 0012 airfoils.}
\end{figure}
\begin{figure}[h!]
    \centering
\includegraphics[width=0.4\linewidth]{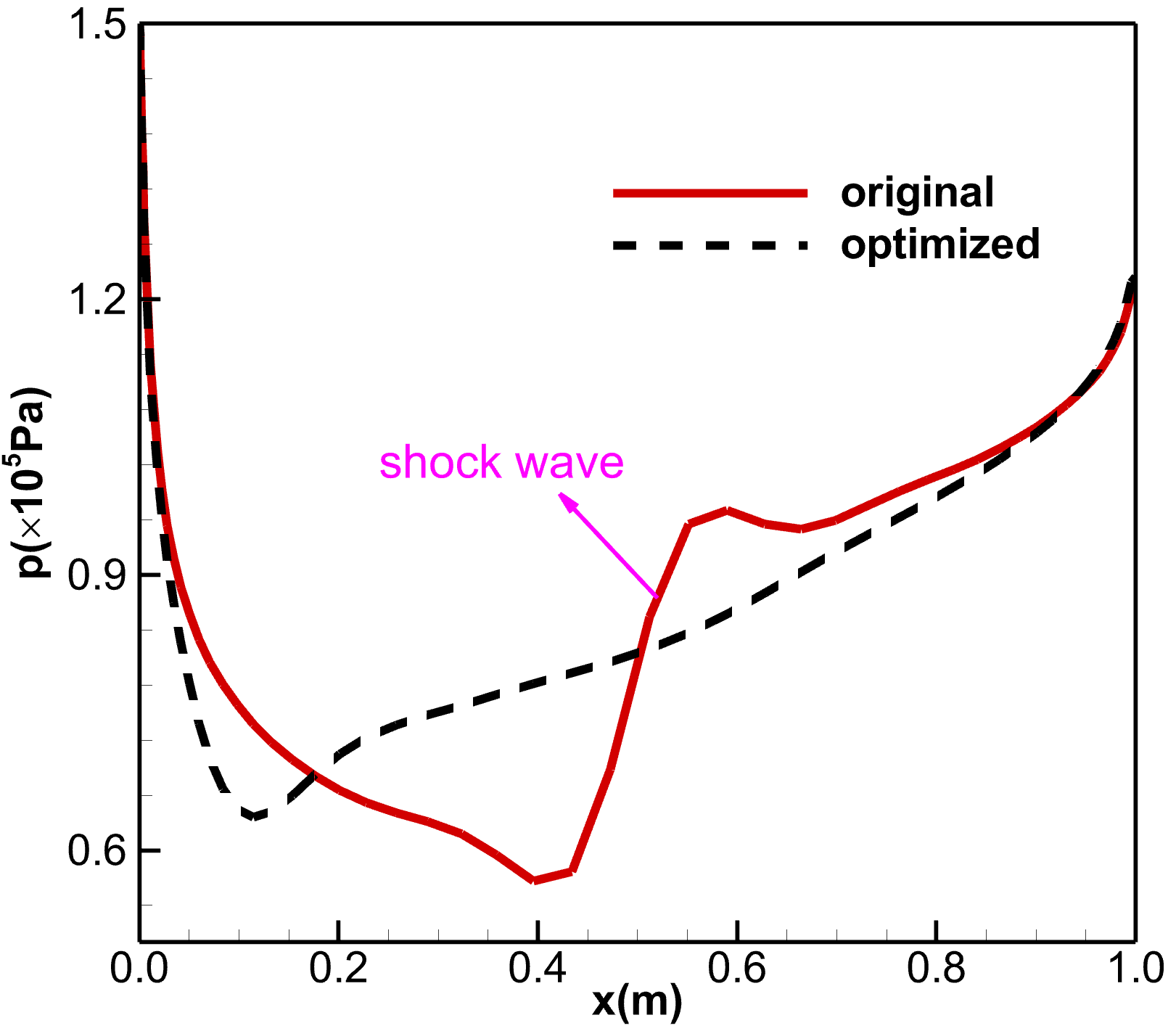}
    \caption{Comparison of the static pressure on the blade surface between the original and optimized NACA 0012 airfoils}
    \label{naca_opt_p_surface2}
\end{figure}

Figures~\ref{naca_Ma_org} and~\ref{naca_Ma_opt} compare the Mach number contours of the original and optimized NACA 0012 airfoils. In the original configuration, a strong shock wave appears on the airfoil surface, with the upstream Mach number reaching approximately 1.2. After optimization, the Mach number upstream of the shock is reduced to about 1.1, resulting in a noticeably weaker shock on the optimized airfoil surface. The distribution of static pressure on the airfoil surfaces, as shown in Fig.~\ref{naca_opt_p_surface2}, further corroborates this behavior.

\begin{figure}[h!]
    \centering
\includegraphics[width=0.4\linewidth]{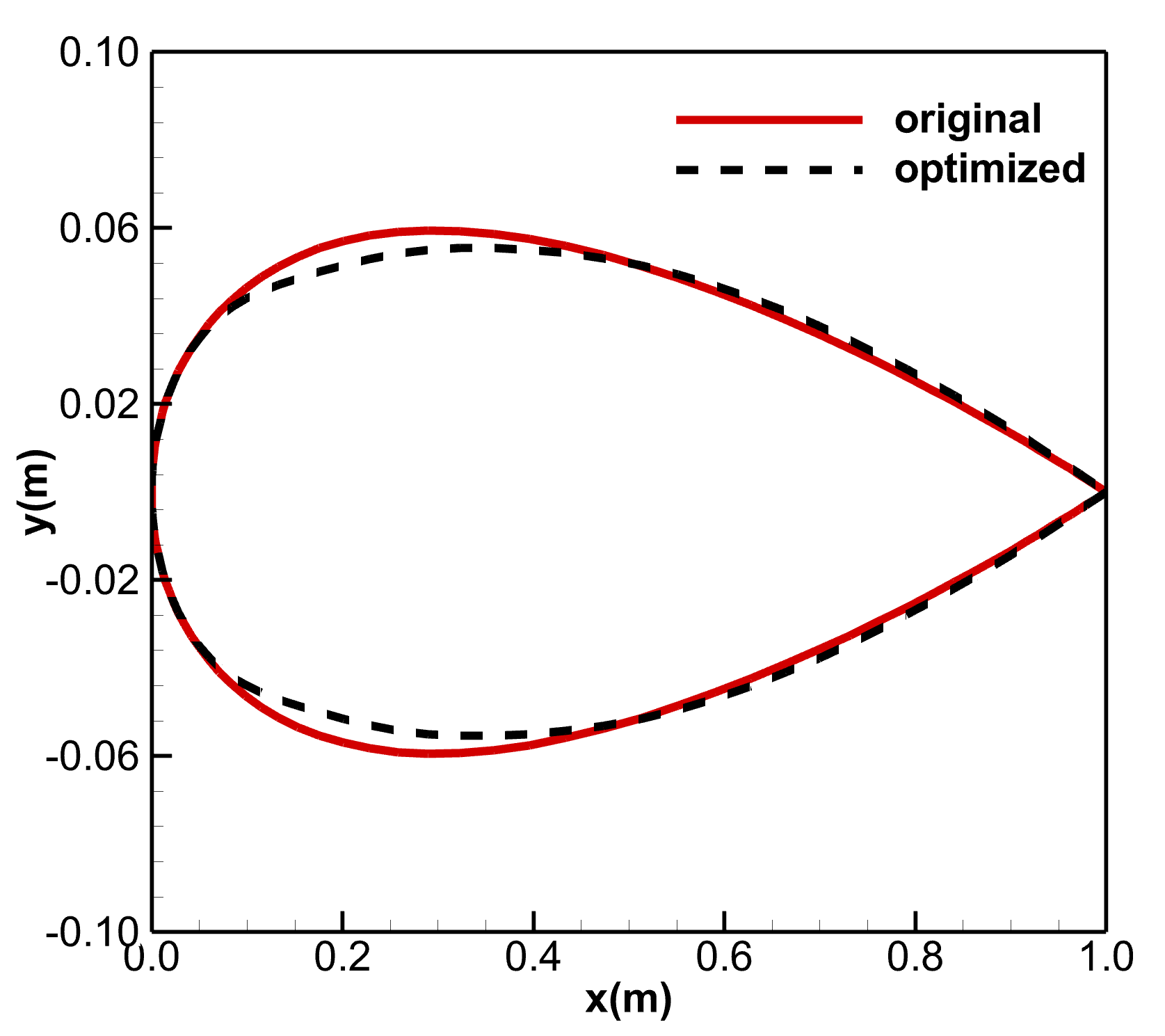}
    \caption{Comparison of the original and optimized NACA 0012 airfoils for the shock-strength reduction optimization}
    \label{naca_airfoil2}
\end{figure}
Figure~\ref{naca_airfoil2} compares the original and optimized airfoils obtained from the shock‑strength–reduction optimization. Relative to the baseline airfoil, the optimized profile shows a reduction in thickness from the leading edge to the mid‑chord, followed by a slight increase toward the trailing region. The reduced thickness near the leading edge decreases the flow acceleration in this region, resulting in a lower upstream Mach number. Since a lower Mach number ahead of the shock leads to a weaker shock wave, the associated entropy generation is consequently diminished.

\section{Conclusion}

In this work, a discrete adjoint gas-kinetic scheme (GKS) is developed for the sensitivity analysis and aerodynamic shape optimization of both external and internal configurations within continuum flow regimes. First, the discrete adjoint solver is rigorously verified from both qualitative and quantitative perspectives. The results indicate that the contours of the adjoint turbulence variables exhibit inverse trends compared to those of the corresponding primal flow variables. Furthermore, the adjoint GKS solver demonstrates sensitivity convergence behaviors and asymptotic root-mean-square (RMS) residual decay rates identical to those of the linearized GKS solver. At full convergence for the Durham turbine cascade case, the relative discrepancies in sensitivity predictions between the adjoint and linearized methods are negligible.
Second, the practical effectiveness of the adjoint GKS-based design optimization framework is demonstrated through three benchmark cases. In the first case—an inverse design problem—the objective function is reduced by $99.9\%$, and the target Mach number distribution is successfully recovered within ten design cycles. The second case focuses on enhancing the lift-to-drag ratio of a NACA 0012 airfoil; the optimized configuration achieves a twofold increase in lift while maintaining a nearly constant drag penalty. The third case addresses shock-strength reduction by minimizing the entropy generation rate. After ten design cycles, the pre-shock Mach number is reduced from 1.2 to 1.1, resulting in a substantially weakened shock wave.

\section*{Acknowledgment}

This work was supported by the National Natural Science Foundation of China (Grant No. 92371107), the National Key R$\&$D Program of China (Grant No. 2022YFA1004500), and the Hong Kong Research Grants Council (Grant No. 16208324).

\section*{Appendix A: Linearized GKS Solver}

\begin{figure}[h!]
    \centering
\includegraphics[width=0.6\linewidth]{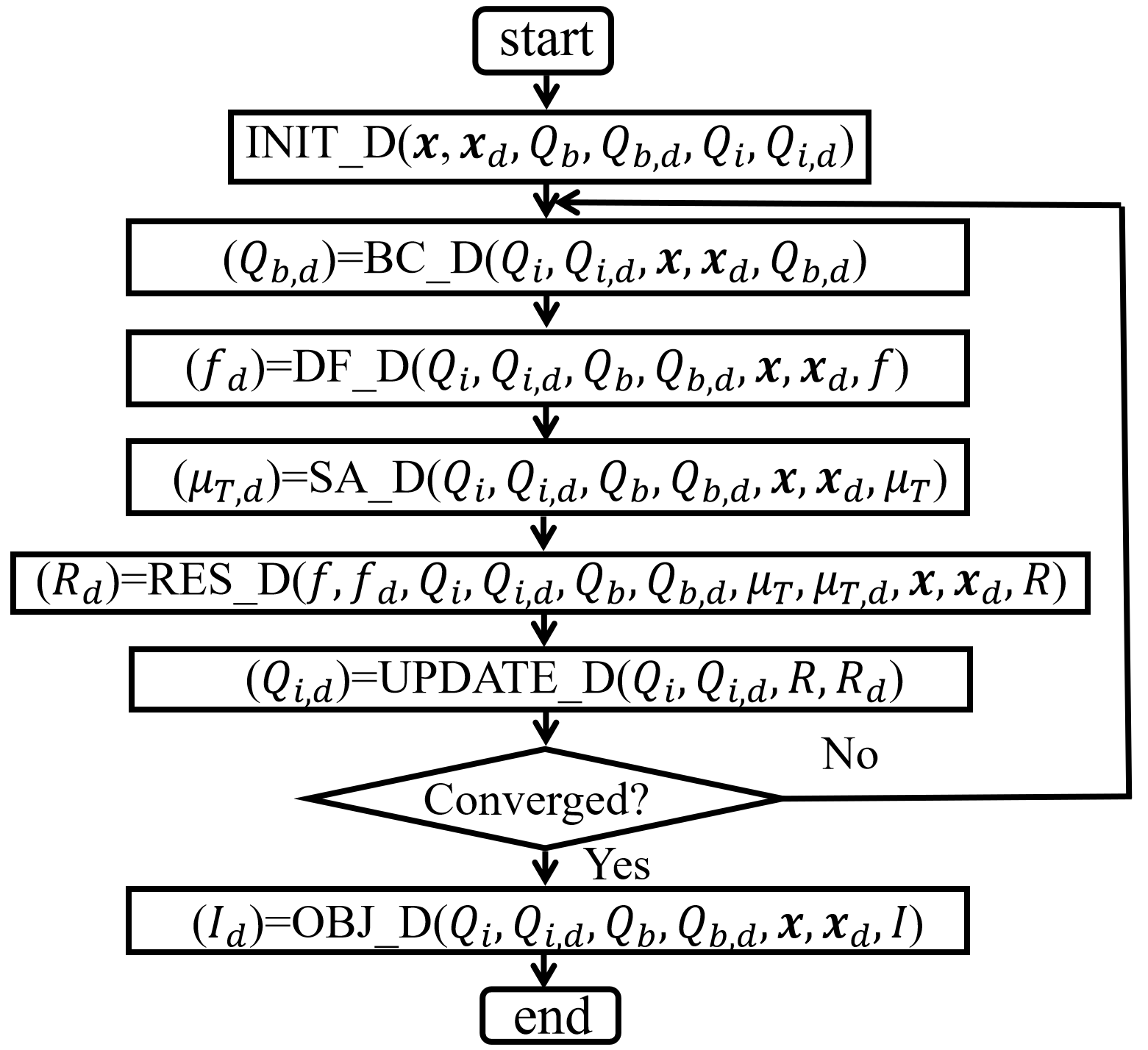}
    \caption{The dataflow of the linearized GKS solver}
    \label{Lin_GKS}
\end{figure}

Figure~\ref{Lin_GKS} illustrates the dataflow of the linearized GKS solver. It can be observed that its dataflow is consistent with that of the flow GKS solver. The linearized solver similarly consists of seven main components, whose functionalities are described below.
\begin{enumerate}
\item INIT\_D: initialize the linearized macroscopic flow variables with 0 and the perturbation of the grid coordinate with 1;
\begin{equation}
Q_{b,d} = 0, Q_{i,d}=0, \boldsymbol x_d = 1
\end{equation}

\item BC\_D: linearize the boundary condition to obtain the linearized macroscopic flow variables at the boundary. According to the chain rule, we have
\begin{equation}
Q_{b,d}=\frac{\partial Q_b}{\partial Q_i}Q_{i,d} + \frac{\partial Q_b}{\partial \boldsymbol x}\boldsymbol x_d
\end{equation}

\item SA\_D: linearize the turbulence model equation to obtain $\mu_{T,d}$, yielding
\begin{equation}
\mu_{T,d} = \frac{\partial \mu_T}{\partial Q_i}Q_{i,d} + \frac{\partial \mu_T}{\partial Q_b}Q_{b,d} + \frac{\partial \mu_T}{\partial \boldsymbol x}\boldsymbol x_d
\end{equation}

\item DF\_D: linearize the distribution function to obtain $f_d$, yielding
\begin{equation}
f_d = \frac{\partial f}{\partial Q_i}Q_{i,d} + \frac{\partial f}{\partial Q_b}Q_{b,d} + \frac{\partial f}{\partial \boldsymbol x}\boldsymbol x_d
\end{equation}

\item RES\_D: linearize the residual to obtain $R_d$, leading to
\begin{equation}
R_d = \frac{\partial R}{\partial f}f_d +\frac{\partial R}{\partial \mu_T}\mu_{T,d}+ \frac{\partial R}{\partial Q_i}Q_{i,d} + \frac{\partial R}{Q_b}Q_{b,d} + \frac{\partial R}{\partial \boldsymbol x}\boldsymbol x_d
\end{equation}

\item UPDATE\_D: update the linearized macroscopic flow variables using the same time-marching scheme as the flow and adjoint GKS solvers, and we have
\begin{equation}
\frac{\partial Q_{i,d}}{\partial t}+R_d =0
\end{equation}

\item OBJ\_D: linearize the objective function to obtain sensitivity information
\begin{equation}
I_d = \frac{\partial I}{\partial Q_i}Q_{i,d} + \frac{\partial I}{\partial Q_b}Q_{b,d} + \frac{\partial I}{\partial \boldsymbol x}\boldsymbol x_d
\end{equation}

\end{enumerate}


\bibliographystyle{elsarticle-num} 
\bibliography{main}

\end{document}